\documentclass[fleqn,usenatbib]{mnras}

\usepackage{mathptmx}

\usepackage[T1]{fontenc}

\DeclareRobustCommand{\VAN}[3]{#2}
\let\VANthebibliography\thebibliography
\def\thebibliography{\DeclareRobustCommand{\VAN}[3]{##3}\VANthebibliography}

\usepackage{graphicx}	
\usepackage{amsmath}	
\usepackage{amssymb}	
\usepackage{soul}
\usepackage{color}

\newcommand{\ec}{ELN-criterion}
\definecolor{myblue}{rgb}{0.8, 0.25, 0.69}

\title[Bar formation criteria in IllustrisTNG]{Disc instability and bar formation: view from the IllustrisTNG simulations}

\author[Izquierdo-Villalba et al.]{David Izquierdo-Villalba,$^{1,2}$\thanks{E-mail: david.izquierdovillalba@unimib.it}
Silvia Bonoli,$^{3,4}$
Yetli Rosas-Guevara,$^{3}$
Volker Springel,$^{5}$
Simon D.M. White,$^{5}$\newauthor
Tommaso Zana,$^{6}$
Massimo Dotti,$^{1,2,7}$
Daniele Spinoso,$^{3,8}$
Matteo Bonetti,$^{1,2,7}$
Alessandro Lupi$^{1,2}$
\\
$^{1}$Dipartimento di Fisica ``G. Occhialini'', Universit\`{a} degli Studi di Milano-Bicocca, Piazza della Scienza 3, I-20126 Milano, Italy\\
$^{2}$INFN, Sezione di Milano-Bicocca, Piazza della Scienza 3, 20126 Milano, Italy.\\
$^{3}$Donostia International Physics Centre (DIPC), Paseo Manuel de Lardizabal 4, 20018 Donostia-San Sebastian, Spain\\
$^{4}$IKERBASQUE, Basque Foundation for Science, E-48013, Bilbao, Spain\\
$^{5}$Max-Planck Institute for Astrophysics, Karl-Schwarzschild-Str. 1, D-85741 Garching, Germany\\
$^{6}$Scuola Normale Superiore, Piazza dei Cavalieri 7, I-56126 Pisa, Italy \\
$^{7}$INAF, Osservatorio Astronomico di Brera, Via E. Bianchi 46, I-23807, Merate, Italy\\
$^{8}$Centro de Estudios de F\'{\i}sica del Cosmos de Arag\'{o}n (CEFCA), Plaza San Juan 1, Planta-2, Teruel, 44001, Spain.\\
}

\date{Accepted XXX. Received YYY; in original form ZZZ}

\pubyear{2022}

\begin{document}
\label{firstpage}
\pagerange{\pageref{firstpage}--\pageref{lastpage}}
\maketitle

\begin{abstract}

\noindent We make use of $z\,{=}\,0$ samples of strongly barred and unbarred disc galaxies from the \texttt{TNG100} and \texttt{TNG50} cosmological hydrodynamical simulations to assess the performance of the  simple disc instability criterion proposed by Efstathiou, Lake \& Negroponte (1982) (\ec{}). We find that strongly barred galaxies generally assemble earlier, are more star-dominated in their central regions, and have more massive and  more compact discs than unbarred galaxies. The \ec{} successfully identifies ${\sim}\,75\%$ and ${\sim}\,80\%$ of the strongly barred and the unbarred galaxies, respectively. Strongly barred galaxies that the criterion fails to identify tend to have more extended discs, higher spin values and bars that assembled later than is typical for the bulk of the barred population. The bars in many of these cases appear to be produced by an interaction with a close neighbour (i.e. to be externally triggered) rather than to result from secular growth in the disc. On the other hand, we find that unbarred galaxies misclassified as barred by the \ec{} typically have stellar discs similar to those of barred galaxies, although more extended in the vertical direction and less star-dominated in their central regions, possibly reflecting later formation times. In addition, the bulge component of these galaxies is significantly more prominent at early times than in the strongly barred sample. Thus, the \ec{} robustly identifies secular bar instabilities in most simulated disc galaxies, but additional environmental criteria are needed to account for interaction-induced bar formation.
\end{abstract}




\section{Introduction}

Nowadays we know that bars are common structures of galaxies in the local Universe where nearly $60$ percent of the disc dominated galaxies host one \citep{Knapen1999,Eskridge2000,Grosbol2004,MenendezDelmestre2007,Barazza2008A}. Bars are believed to be important for the secular evolution of disc-dominated galaxies. Their capability of redistributing the galaxy angular momentum leads to gas inflows towards the galaxy nuclei, triggering star formation episodes which ultimately can lead to the formation of a pseudobulge \citep{LyndenBellKalnajs1972,Tremaine1984,Kormendy1993,vanAlbada1981,Schwarz1981,Sakamoto1999}. Even though current simulations are able to track the evolution of stellar discs \citep{Navarro1991,Navarro2000}, it remains unclear why certain galaxies end up developing a bar structure whereas other similar galaxies do not. One of the pioneering works trying to shed light on the physical conditions that lead to bar formation was  \cite{Efstathio1982}, which explored the global stability of cold exponential stellar discs by performing a set of 2D N-body simulations. Interestingly, the authors found a simple analytical definition  to determine the stability against bar formation:
\begin{equation} \label{eq:Bar_formation_Efstathiou1982}
\rm  \epsilon\,{=}\,\frac{\mathit{V}_{max}}{(\mathit{G} \mathit{M}_{disc}/\mathit{R}_{\rm d})^{1/2}} 
\end{equation}
where $G$ is the gravitational constant, $\rm \mathit{V}_{max}$ is the maximum rotational velocity of the system, $R_{\rm d}$ is the scale length of the stellar disc and $\rm \mathit{M}_{disc}$ its total mass. The simulations showed that galaxies with $\epsilon \, {>}\, 1.1$ possess a large enough hot component able to stabilize the stellar disc. Instead, galaxies with  $\epsilon\,{\leq}\,1.1$ \, have stellar discs that become bar unstable. Thanks to the simplicity of the criterion (\ec{} hereafter), years later \cite{Mo1998} adopted it to build a phenomenological model of disc galaxy formation within a hierarchical galaxy formation model.\\

Subsequent works addressing the bar formation process focused on the role of the dark matter (DM) component. Unlike \cite{Efstathio1982}, they included \textit{live} DM halos that are capable to interact dynamically with the stellar structure \citep[see e.g][]{Debattista1998,HolleyBockelmann2005,Weinberg2007a,Weinberg2007b,RomanoDiaz2008,Dubinski2009,Saha2012}.  \cite{Athanassoula2002} compared two numerical simulations of isolated galaxies with the same disc-to-halo ratio, showing that the galaxy with larger halo concentration developed a much stronger, larger and thinner bar. Besides, \cite{Debattista1998}, \cite{Debattista2000} and \cite{Athanassoula2003} showed that halo concentration can leave an imprint on the bar strength and pattern speed. On top of concentration, the total halo mass in which a galaxy reside might play an important role as well. By performing numerical simulations, \cite{Athanassoula2003} showed that while the most massive halos hosted the systems with the strongest bars, the less massive ones displayed weaker bar structures. During the last years, many studies have focused also on the effect of the halo spin in the stability of the stellar disc against bar modes. \cite{Saha2013b} reported that dark matter halos with a spin parameter between $0$ and $0.07$ in co-rotation with the stellar disc are able to prompt the formation of bars and boxy bulges. \cite{Long2014} found the opposite trend, showing that bar formation in spinning dark matter halos might be heavily suppressed. Along the same line, \cite{Collier2018} showed that bars hosted in halos with the largest spin had difficulties in re-growing bars after a buckling instability. Even more, in many cases bars were dissolved, leaving behind a host disc with large radial dispersion velocities. The authors attributed this bar dumping to the difficulty of spinning DM halos to absorb additional angular momentum.\\

On top of the dark matter component, the bulge one can play a major role in the bar assembly and evolution \citep{Ostriker1973}. For instance, \cite{Kataria2018} using N-body simulations of isolated galaxies, showed a delay in the bar formation as a function of the galaxy bulge-to-disc ratio (B/D). Particularly, they reported a $\rm B/D$ upper cut off above which the development of a bar is suppressed. While in dense bulges this cut was at $\rm B/D\,{\sim}\,0.2$, in the less dense ones was at $\rm B/D\,{\sim}\,0.5$ (see also the work of \citealt{Kataria2020}). Similar dependencies with the bulge prominence could be derived by the comparison of the twins zoom-in hydrodynamical simulations \texttt{Eris} and \texttt{ErisBH} \citep{Guedes2011,Bonoli2016}, where the bar structure developing only in \texttt{ErisBH} is likely favoured by the smaller bulge of this galaxy with respect to the one of \texttt{Eris} \citep{spinoso2017}. Some observational studies have also shown a correspondence between the presence of bulges and bars. For instance, the results of \cite{Barazza2008} and \cite{Aguerri2009} pointed out that the bar fraction decreases with increasing bulge luminosity.\\

During the last years, semi-analytical models (SAMs) have tried to shed light on the evolution of barred galaxies inside a hierarchical universe. For that, they have generally relied on the aforementioned \ec{}, which has the advantage of being simple and depending only on global galaxy properties, easily accessible in SAMs \citep{Guo2011,Barausse2012,Croton2016,Lacey2016,Cora2018}. Although the \ec{} does not take into account all the dependencies discussed above \citep[see the discussion of][]{Athanassoula2008}, SAM predictions display, in general, a good agreement with observations \citep[see for instance][]{IzquierdoVillalba2019,Irodotou2019,Marshall2019a}. Despite its important role in this context, no systematic study about the performance and reliability of the \ec{} in a cosmological context has done to date. Some attempts in this direction can be found in \cite{Yurin2015}, who explored the performance of the \ec{} by using a methodology to include live stellar discs into high-resolution Milky Way-like dark matter halos. Interestingly, the authors reported that the \ec{} should be taken as an important first guide to discriminate between bar stable or unstable discs. Simulations of isolated dark matter halos with an embedded stellar disc performed by \cite{Mayer2004} pointed towards the same direction. Even more, the authors suggested that the disc-to-halo ratio within the typical disc radius might be the main factor determining the final development of barred structures \citep[see similar results of][]{DeBuhr2012}. Even though these works are important for supporting the \ec{} as a necessary condition for bar formation, they could not test its performance with a large galaxy sample, evolving consistently in their full cosmological context. Such tests are now possible thanks to the latest generation of cosmological hydrodynamical simulations, which can currently follow the physical assembly of galaxies down to relatively small scales in representative cosmological volumes \citep[see e.g,][]{Dubois2014,Schaye2015,Nelson2018,Dave2019}. Indeed, it has been shown that current  cosmological hydrodynamical simulations are capable to reproduce the barred galaxy population at $z\,{=}\,0$ (see \citealt{Algorry2017}, \texttt{EAGLE}, \citealt{PeschkenLokas2019}, \texttt{Illustris}, \citealt{RosasGuevara2019,RosasGuevara2021}, \texttt{IllustrisTNG}, \citealt{Reddish2021}, \texttt{NewHorizonAGN}, \citealt{Fragkoudi2020}, \texttt{AURIGA}), even though some tensions still remain \citep[see e.g,][]{Fragkoudi2020,Roshan2021}.\\

In this paper we systematically explore, for the first time, the performance of the \cite{Efstathio1982} analytical criterion using a sample of barred and unbarred galaxies extracted from a large cosmological hydrodynamical simulation. In particular, we focus on  $z\,{=}\,0$ Milky-Way type galaxies ($\rm 10^{10.4} \, M_{\odot}\, {\lesssim}\,M_{stellar}\,{\lesssim}\,10^{11}\, M_{\odot}$) extracted from the \texttt{TNG100} and \texttt{TNG50} simulations. The outline of this work is as follows. In Section~\ref{sec:MEtodology} we describe the main characteristics of the \texttt{TNG100}/\texttt{TNG50} simulations and the barred/unbarred galaxy sample. Besides, we present our methodology to extract the stellar disc scale length. In Section~\ref{sec:AssemblyBarsUnbars} we analyse the general properties of barred and unbarred galaxies. In Section~\ref{sec:ECriterionPerformance} we test the success rate of the \ec{}, we study the characteristics of correctly and incorrectly classified barred/unbarred galaxies. Finally, in Section~\ref{sec:Conclusions} we summarize our main findings. A Lambda Cold Dark Matter $(\Lambda$CDM) cosmology with parameters $\Omega_{\rm m} \,{=}\,0.309$, $\Omega_{\rm \Lambda}\,{=}\,0.691$, $\Omega_{\rm b}\,{=}\,0.047$, $\sigma_{8}\,{=}\,0.816$ and $\rm H_0\,{=}\,67.74\, \rm km\,s^{-1}\,Mpc^{-1}$ is adopted throughout the paper \citep{PlanckCollaboration2016}. Unless explicitly stated, all the distances used in this work are physical distances.

\begin{figure*}
\centering
\includegraphics[width=1\columnwidth]{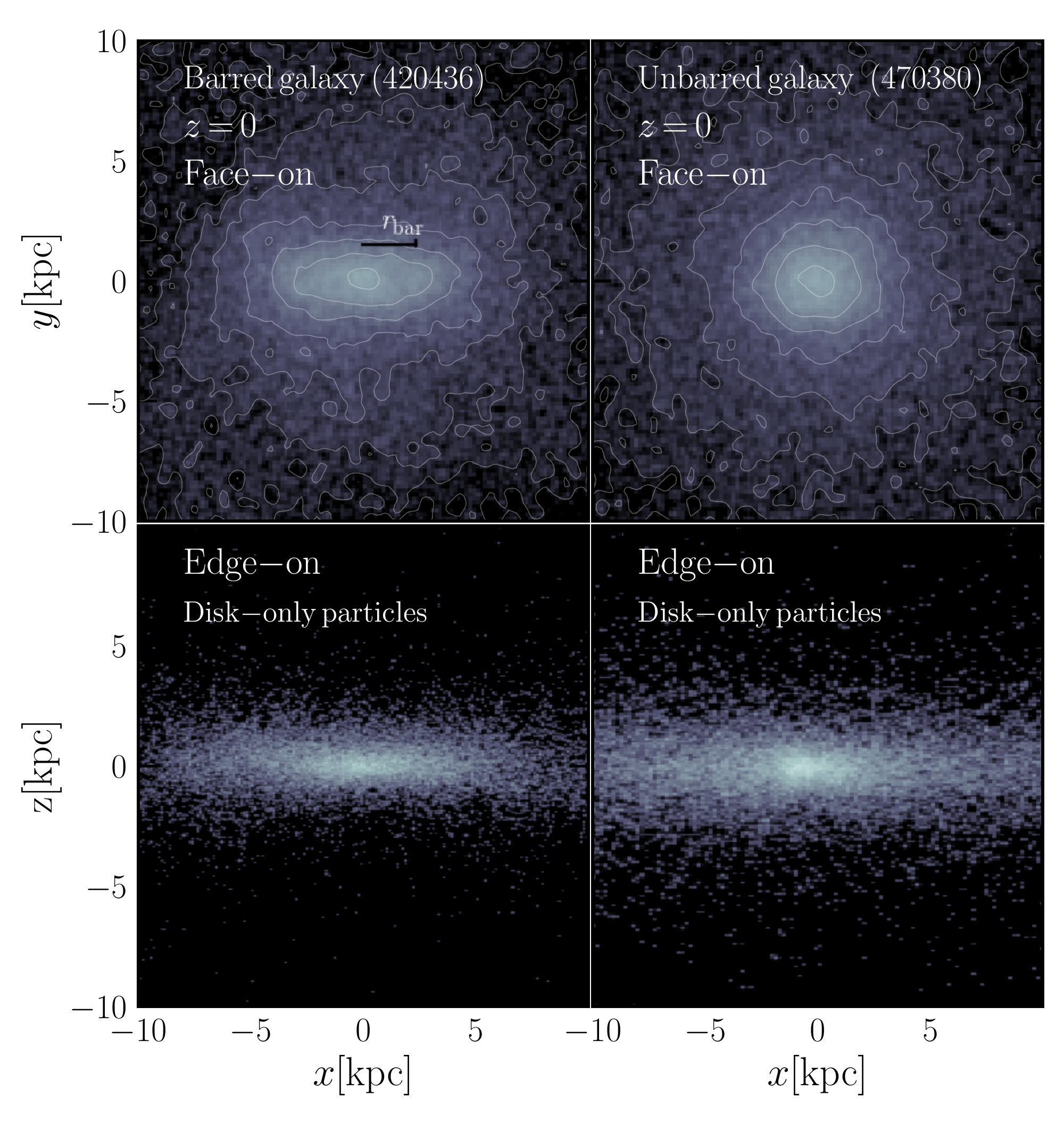}
\includegraphics[width=1\columnwidth]{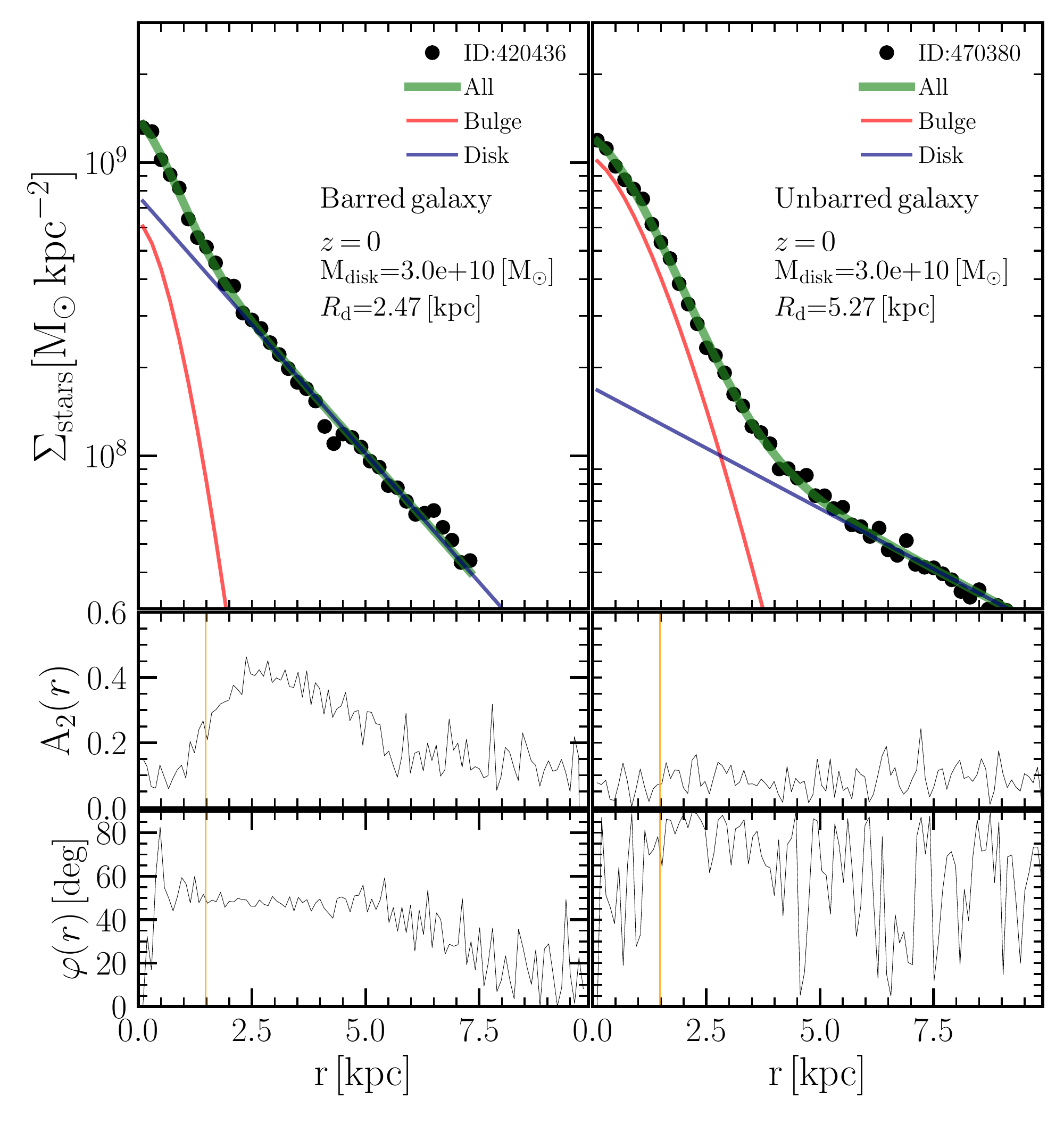}
\caption{\textbf{Left panel}: Galaxy face-on (up) and edge-on (down) surface mass density maps for a \texttt{TNG100} barred (left) and \texttt{TNG100} unbarred (right) galaxy at $z\,{=}\,0$. For the edge-on view we only show the disc particles (see Eq.~\ref{eq:Circularity}). \textbf{Right panel}:  The upper panels display the galaxies (barred left and unbarred right) face-on surface density profile: pink, red and green represent, respectively, the fit of the whole galaxy, the bulge and the disc component. The middle and lower panels show the value of $\mathrm{A_2}(r)$ and  $\varphi(r)$, respectively. To guide the reader, the vertical orange lines highlight the position at which $r$ is equal to twice the softening length of \texttt{TNG100} at $z\,{=}\,0$.}
\label{fig:Example_descomposition_Bulge_disc}
\end{figure*}

\section{The galaxy sample} \label{sec:MEtodology} 

In this section we present our sample of simulated barred/unbarred galaxies, extracted from the \textit{The Next Generation Illustris Simulations}\footnote{The simulations of the \texttt{TNG} project are available
at \href{https://www.tng-project.org/}{https://www.tng-project.org/} \citep{NelsonTNGDataReleas2019}} (hereafter, \texttt{TNG}, \citealt{NelsonTNGDataReleas2019}). Specifically, we make use of \texttt{TNG100} and \texttt{TNG50} whose volume, mass and spatial resolution guarantees a good sampling of massive disc-dominated galaxies \citep[see e.g,][]{RosasGuevara2019,Du2019,Zhou2020,RosasGuevara2021,Gargiulo2021}. Among the entire galaxy population, we focus on $z\,{=}\,0$ Milky-Way type galaxies ($\rm 10^{10.4} {\lesssim}\,M_{stellar}\,{\lesssim}\,10^{11}\, M_{\odot}$ and disc-to-total ratio, $\rm D/T\,{>}\,0.5$) which have proved to be the preferential hosts and birthplaces of bar structures in the low-$z$ Universe \citep[see e.g][]{Gadotti2009,Cervantes2015,Gavazzi2015,IzquierdoVillalba2019,Marshall2019a}.\\

\subsection{\texttt{IllustrisTNG} project}

The \texttt{TNG} project is a set of simulations of large cosmological volumes run with the moving-mesh code \texttt{AREPO} \citep{Springel2010}. To follow the evolution of galaxies and subhalos, \texttt{AREPO} solves the magneto-hydrodynamics equations coupled with self-gravity and includes an updated version of the \texttt{Illustris}\footnote{\href{https://www.illustris-project.org/}{https://www.illustris-project.org/}} \textit{subgrid} galaxy physics \citep{Vogelsberger2014a,Vogelsberger2014b}. The \textit{subgrid} model includes prescriptions for radiative gas cooling, stellar evolution, AGN/supernovae feedback, chemical enrichment, gas recycling, black hole seeding, black hole growth and metal loading of outflows \citep[see further details in][]{Weinberger2017,Pillepich2018b}.\\ 

\noindent The \texttt{TNG} project is made up of three different simulation volumes (\texttt{TNG300}, \texttt{TNG100} and \texttt{TNG50}) evolved from $z\,{=}\,127$ down to $z\,{=}\,0$ with cosmological parameters from \cite{PlanckCollaboration2016}. 
Among the three volumes, \texttt{TNG100}  \citep[][]{Pillepich2018a,Springel2018,Nelson2018,Naiman2018,Marinacci2018}  follows in a $75\,\mathrm{Mpc}/h$ comoving box size the evolution of $2\,{\times}\,1820^3$ dark matter particles and gas cells with mass of $\rm 7.46\,{\times}\,10^6\,M_{\odot}$ and $\rm 1.39\,{\times}\, 10^6 M_{\odot}$, respectively. At $z\,{=}\,0$, the softening length of collisionless particles and gas component corresponds, respectively, to $\rm 0.5\, kpc/\mathit{h}$ and $\rm 125\,pc/\mathit{h}$. On the other hand, \texttt{TNG50} \citep{Nelson2019,Pillepich2019} simulates a smaller volume ($\rm 35 Mpc/\mathit{h}$ box size) but follows dark matter and gas using $2160^3$ elements each, with masses of $\rm 4.5\,{\times}\,10^5 \, M_{\odot}$ and $\rm 8.5\,{\times}\,10^4 \, M_{\odot}$ . By $z\,{=}\,0$, the softening length of collisionless particles and gas component corresponds to $\rm 0.195\, kpc/\mathit{h}$ and $\rm 50\,pc/\mathit{h}$, respectively. Particle data of both \texttt{TNG100} and \texttt{TNG50} were stored in 100 different snapshots from $z\,{\sim}\,20$ to $z\,{=}\,0$.  Dark matter subhalos and galaxies were identified within these snapshots by using a friend-of-friend group finder \citep{Davis1985} and the \texttt{SUBFIND} algorithm \citep{Springel2001}. Finally, by applying \texttt{L-HALOTREE} and \texttt{SUBLINK} algorithms \citep{Springel2005,RodriguezGomez2015} all DM subhalos and galaxies were arranged in merger tree structures.

\subsection{Bar and unbarred galaxies in \texttt{TNG100} and \texttt{TNG50}} \label{sec:Bars_NoBars_and_Fit}

In this work we make use of the $z\,{=}\,0$ \texttt{TNG100}/\texttt{TNG50} barred and unbarred galaxy sample of \cite{RosasGuevara2019} 
and \cite{RosasGuevara2021}. For the \texttt{TNG100} simulation, \cite{RosasGuevara2019} selected a $z\,{=}\,0$ galaxy sample with \textit{disc-to-total} ratio ($\rm D/T$) ${>}\,0.5$, guaranteeing morphologies fully dominated by the disc structure. To ensure a well-resolved disc, the extra condition of more than $10^4$ stellar particles within twice the galaxy half mass radius ($R_{\rm half}$) was also imposed. Such a particle limit sets a lower stellar mass cut of $\rm M_{stellar}\,{\sim}\,10^{10.4}\, M_{\odot}$. From this disc-dominated galaxy sample, the authors identified stellar bars via a Fourier decomposition of the face-on stellar surface density ($\Sigma_{\rm stars}$, \citealt{Athanassoula2002,Valenzuela2003}). Specifically, they determined the strength of a non-axisymmetric nuclear component ($\rm A_2$) by computing the ratio between the second and zero terms of the Fourier expansion:

\begin{equation}\label{eq:A2}
\rm A_2(\mathit{r}) \,{=}\, \frac{\left| \sum_{j} M_{\mathit{j}}\, e^{2\mathit{i} \theta_{\mathit{j}}}\right|}{\sum_{j} M_{\mathit{j}}} ,
\end{equation}
where $\rm M_{\mathit{j}}$, $\theta_{j}$, and $r$ are the mass, angular coordinate in the galactic plane and radial distance of the ${j{-}{\rm th}}$ stellar particle, respectively. The sums of Eq.~\ref{eq:A2} were performed over all particles within cylindrical shells of $0.12 \rm \, kpc$ width and $\rm 2 \, kpc$ height. In short, $\rm A_2(\mathit{r})$ displays an increasing trend up to a distance $r_{\rm max}$ where the strength of the non-axisymmetric structure exhibits a maximum ($\rm A_2^{max}$). After  $r_{\rm max}$, the values of $\rm A_2(\mathit{r})$ gradually decrease to zero. The position $r_{\rm max}$ was used as an estimate of the bar length ($r_{\rm bar}$). Finally, to avoid confusion between bars and other non-axisymmetric structures such as spiral arms (or spurious detection), the authors imposed that the phase of the second Fourier mode \citep[see][]{Zana2019}:
\begin{equation}
\varphi(r)\,{=}\, \rm \frac{1}{2} \arctan\left[ \frac{\sum_{j} M_{\mathit{j}}\, \sin\left( 2\theta_\mathit{j} \right)}{\sum_{j} M_{\mathit{j}}\, \cos\left( 2\theta_\mathit{j} \right)} \right] 
\end{equation}
\noindent has to be constant within the bar length. This Fourier decomposition allowed \cite{RosasGuevara2019} to divide the disc dominated galaxy sample in three different groups: \textit{unbarred sample} ($\rm A_2^{max}\,{<}\,0.2$), \textit{weak bar sample} ($\rm 0.2\,{\leq}\,A_2^{max}\,{<}\,0.3$) and \textit{strong bar sample} ($\rm A_2^{max}\,{\geq}\,0.3$). As an example, in Fig.~\ref{fig:Example_descomposition_Bulge_disc} we show the $\rm A_2(\mathit{r})$ and $\varphi(r)$ profiles for a strong barred and unbarred galaxy in \texttt{TNG100}. In what follows, among all the \texttt{TNG100} galaxies selected by \cite{RosasGuevara2019}, we only study the unbarred sample and strong bars (hereafter \textit{barred sample} or \textit{barred galaxies}). As reference, the total number of barred and unbarred galaxies is respectively 58 and 131, spanning a range in stellar masses between $\rm 10^{10.4} \, \rm M_{\odot}$ and $10^{11} \, \rm M_{\odot}$\footnote{The parent sample of disc-dominated galaxies in the $\rm 10^{10.4}{<}M_{stellar}{<}10^{11} \,M_{\odot}$ mass range satisfying the previous cuts contains 270 galaxies. We refer to \cite{RosasGuevara2019} for further details.}.\\

To extract a disc galaxy sample in \texttt{TNG50}, \cite{RosasGuevara2021} applied a similar methodology to the one used for \texttt{TNG100}. Given the better mass resolution, the authors extended their analysis down to $\rm M_{stellar}\,{=}\,10^{10} M_{\odot}$, finding $349$ disc dominated galaxies at $z\,{=}\,0$ of which $105$ displayed $\rm A_2^{max}\,{>}\,0.2$. Since our purpose is to compare \texttt{TNG100} and \texttt{TNG50} predictions and to explore how they vary with the resolution of the simulation, among all the disc galaxies of \cite{RosasGuevara2021}, we only select the ones with $\rm M_{stellar}({<}2\mathit{R}_{\rm half})\,{>}10^{10.4} \, M_{\odot}$ and $\rm D/T\,{>}\,0.5$. For the selection of the barred sample, an extra cut of $\rm A_2^{max}\,{\geq}\,0.3$ was performed. The final catalogue of \texttt{TNG50} contains respectively $39$ and $31$ barred and unbarred galaxies. As we can see, the ratio between the number of barred and unbarred galaxies is smaller in \texttt{TNG100} than \texttt{TNG50}. Besides cosmic variance, this difference could be caused by the fact that \texttt{TNG50} simulation finds a larger number of disc-dominated galaxies than the former one. Nevertheless, exploring such differences in detail goes beyond the scope of this paper.\\

\begin{figure}
\centering
\includegraphics[width=1\columnwidth]{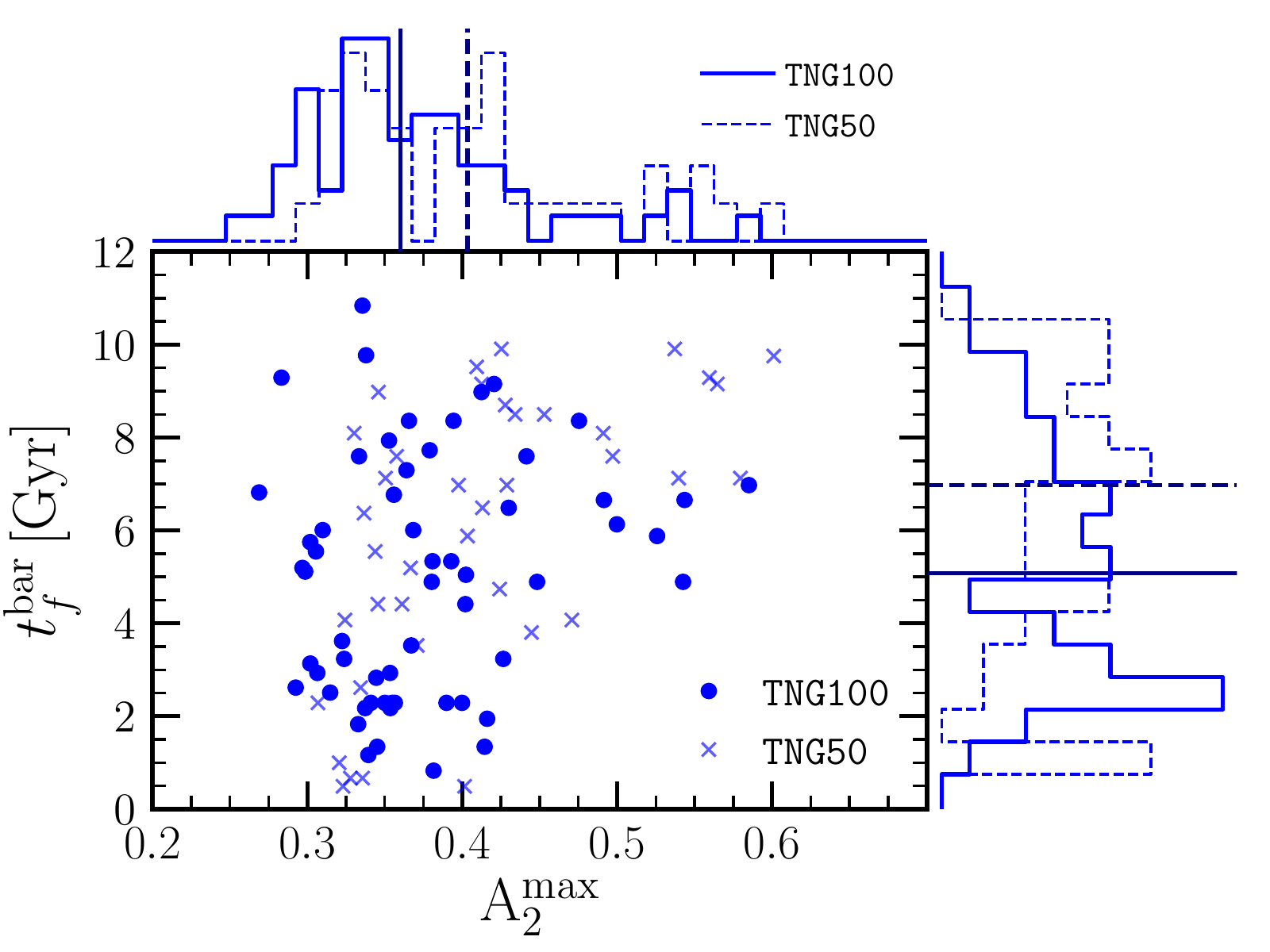}
\caption[]{\textbf{Left panels}: Relation between the bar strength ($\rm A_2^{max}$) at $z\,{=}\,0$ and the bar formation time ($t_f^{\rm bar}$) for \texttt{TNG100} and \texttt{TNG50} (circles and crosses, respectively). Upper and right histograms represent the distribution of $\rm A_2^{max}$ and $t_{f}^{\rm bar}$ for \texttt{TNG100} (solid lines) and \texttt{TNG50} (dashed lines). Vertical lines display the median of the distributions.}
\label{fig:A2_tlookbacktime}
\end{figure}

Following \cite{RosasGuevara2019}, in this work we define the bar formation (loockback) time, $t_{f}^{\rm bar}$ (and its corresponding redshift, $z_f^{\rm bar}$), as the moment in which:
\begin{equation}
    \mathrm{A_2^{max}}(t_{f}^{\rm bar})\,{>}\,0.2   \, \, \, \, \, \,  \& \, \, \, \, \, \, \frac{|\mathrm{A_2^{max}}(t_{f}^{\rm bar}) \, {-} \, \mathrm{A_2^{max}}(t)|} {\mathrm{A_2^{max}}(t_{f}^{\rm bar})} \,{<}\, 0.4 \, ,
\end{equation}
\noindent where $\mathrm{A_2^{max}}(t)$ is the bar strength at the two simulation outputs before the bar formation. For unbarred galaxies we set their $t_{f}^{\rm bar}$ to the median value of the $t_{f}^{\rm bar}$ measured for the whole barred sample. Given that unbarred galaxies never develop a bar, this quantity does not have a physical meaning but it will serve us as a reference to compare with the barred sample. Finally, we also define a normalised time since bar formation, $\delta t$, computed as:
\begin{equation} \label{eq:Normalizet_time_bar}
\delta t \,{=}\,\frac{t_{f}^{\rm bar} \,{-}\, t_{\rm snp}}{t_{f}^{\rm bar}},
\end{equation}
where $t_{\rm snp}$ corresponds to the loockback time of the simulation snapshot. $\delta t \,{=}\,0$ represents the bar formation time while $\delta t \,{>}\,0$ ($\delta t \,{<}\,0$) corresponds to times after (before) the bar formation. In Fig.~\ref{fig:A2_tlookbacktime} we present $\rm A_2^{max}$ and $t_f^{\rm bar}$ at $z\,{=}\,0$ for barred galaxies in \texttt{TNG100} and \texttt{TNG50}. As we can see, bar structures are slightly stronger and older in \texttt{TNG50}, which might be possibly due to resolution effects \citep{Frankel2022}. Regardless of these differences, we can see a weak trend of smaller bar strength at smaller $t_f^{\rm bar}$. Besides, as reported by \cite{RosasGuevara2019}, we find that the larger is the mass of the galaxy, the older is the bar structure (see Figure 6 of \citealt{RosasGuevara2019}).

\subsubsection{Characterization of the disc properties} \label{sec:VerticalScaleLength}

Given that the \ec{} depends on the stellar disc mass and its scale length (see Eq.\ref{eq:Bar_formation_Efstathiou1982}), we need to extract these quantities from our sample of disc-dominated galaxies. To do that, for each galaxy we calculate the angular momentum of the baryons\footnote{We select only gas and stellar particles within $20\,\rm kpc$ around the most bound particle of the galaxy.} ($\vec{J}_{\rm baryons}$) and rotate its reference system in such a way that we align $\vec{J}_{\rm baryons}$ with the z-axis. After the rotation, we compute the galaxy stellar face-on surface density profile, $\Sigma_{\rm stars}(r)$ and we fit it to a composite model obtained as the sum of a Sérsic model \citep{Sersic1968} and an exponential profile:
\begin{equation}\label{eq:Double_sersic_profile}
\Sigma_{\rm stars}(r)\,{=}\,\Sigma_{\rm b}\,e^{-b_n\left[ \left( \frac{r}{\rm \mathit{R}_{b}}\right)^{\frac{1}{n}}  - 1\right] } \, + \, \Sigma_{0}^{\rm d} e^{-\frac{r}{\mathit{R}_{\rm d}}},
\end{equation} 
where the first and second term represent the bulge and disc component. $\Sigma_{\rm b}$, $\rm \mathit{R}_{b}$ and $n$ correspond to the central surface density, effective radius and Sérsic index of the bulge component. On the other hand, $\Sigma_{0}^{\rm d}$ and $\rm \mathit{R}_{\rm d}$ are the central surface density and the scale length of the galaxy stellar disc. The $b_n$ value is such as $\Gamma(2n)\,{=}\,2\gamma(2n,b_n)$ where $\Gamma$ and $\gamma$ are, respectively, the \textit{complete} and \textit{incomplete} gamma function. The fits have been done by finding first the optimal parameters of the exponential disc profile and afterwards we model with the Sérsic profile the central residual excess, i.e the bulge \citep[see the same approach performed in][]{Scannapieco2011,Marinacci2014}. We have used the kinematic bulge-to-disc decomposition of \cite{Genel2015} as an initial guess for the break at which the change between the exponential and the Sérsic law happens. These fits have been done up to $2R_{\rm half}$ and throughout the whole cosmological evolution of the galaxy. In Appendix~\ref{appendix:Check_Galaxy_Decomposition} we show that our $\rm \mathit{R}_{\rm d}$ values are consistent with the ones obtained by only fitting the surface mass density of the thin disc structure, selected according to the kinematic decomposition of Zana et al. (submitted). As an example of the fit performance, in Fig.~\ref{fig:Example_descomposition_Bulge_disc} we present the face-on stellar surface density profile and its disc-bulge decomposition for a barred and unbarred galaxy at $z\,{=}\,0$ in \texttt{TNG100}. As we can see, in both galaxies the structural decomposition finds a well behaved exponential declining trend corresponding to the galactic disc structure.\\

\begin{figure}
\centering\includegraphics[width=1.0\columnwidth]{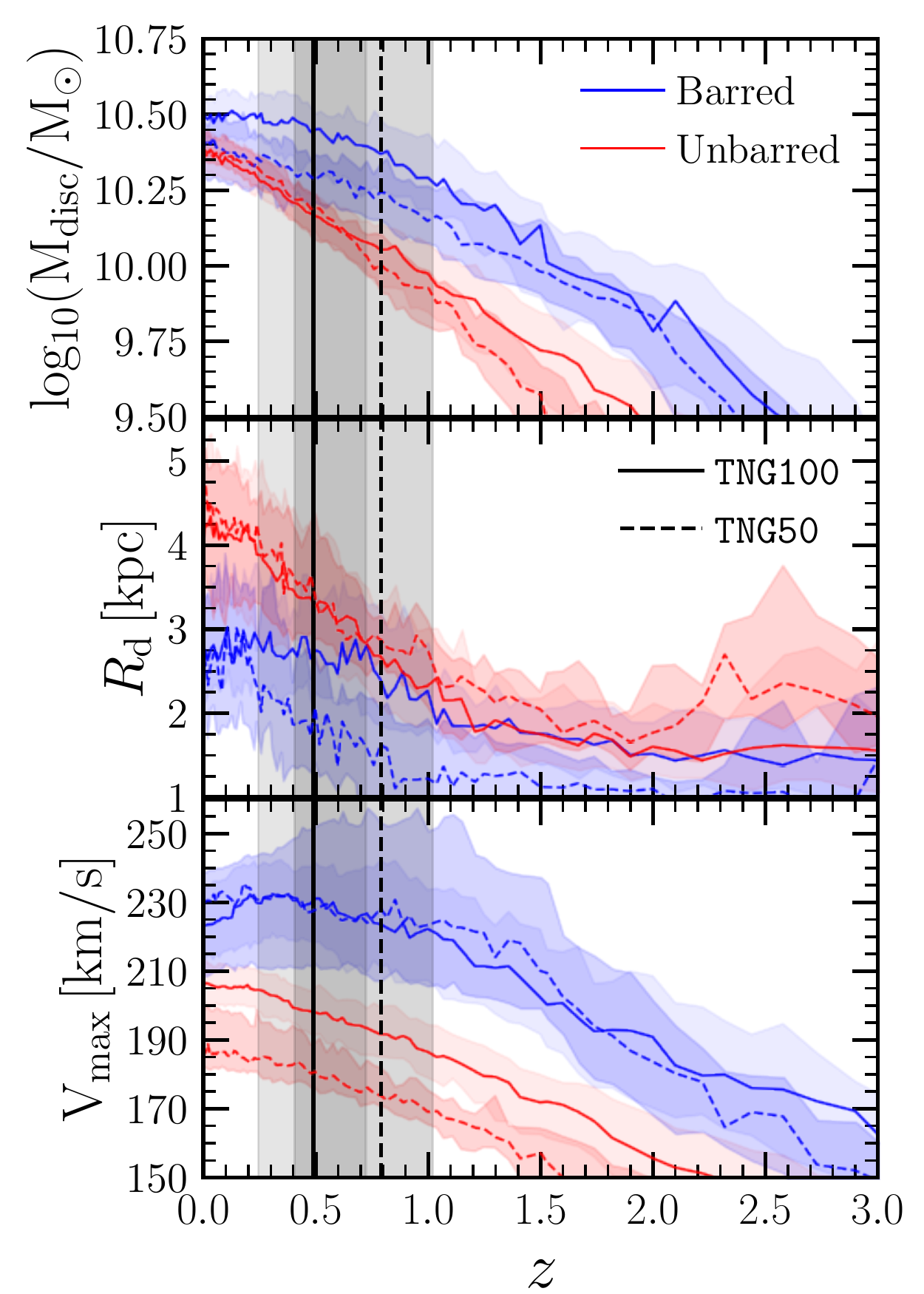}
\caption[]{Redshift evolution of the stellar disc ($\rm M_{disc}$), stellar scale length radius ($R_{\rm d}$) and maximum circular velocity  ($\rm V_{max}$) for our $z\,{=}\,0$ barred (blue) and unbarred (red) sample. While solid (dashed) lines display the median value for \texttt{TNG100} (\texttt{TNG50}), the shaded areas give the $\rm 32^{th}\,{-}\,68^{th}$ percentiles. The solid (dashed) vertical line highlights the median redshift of the bar formation in \texttt{TNG100} (\texttt{TNG50}) and the shaded grey area its $\rm 32^{th}\,{-}\,68^{th}$ percentile.}
\label{fig:Mdisc_Rdisc_Vmax_Bars}
\end{figure}

\begin{figure}
\centering
\includegraphics[width=1\columnwidth]{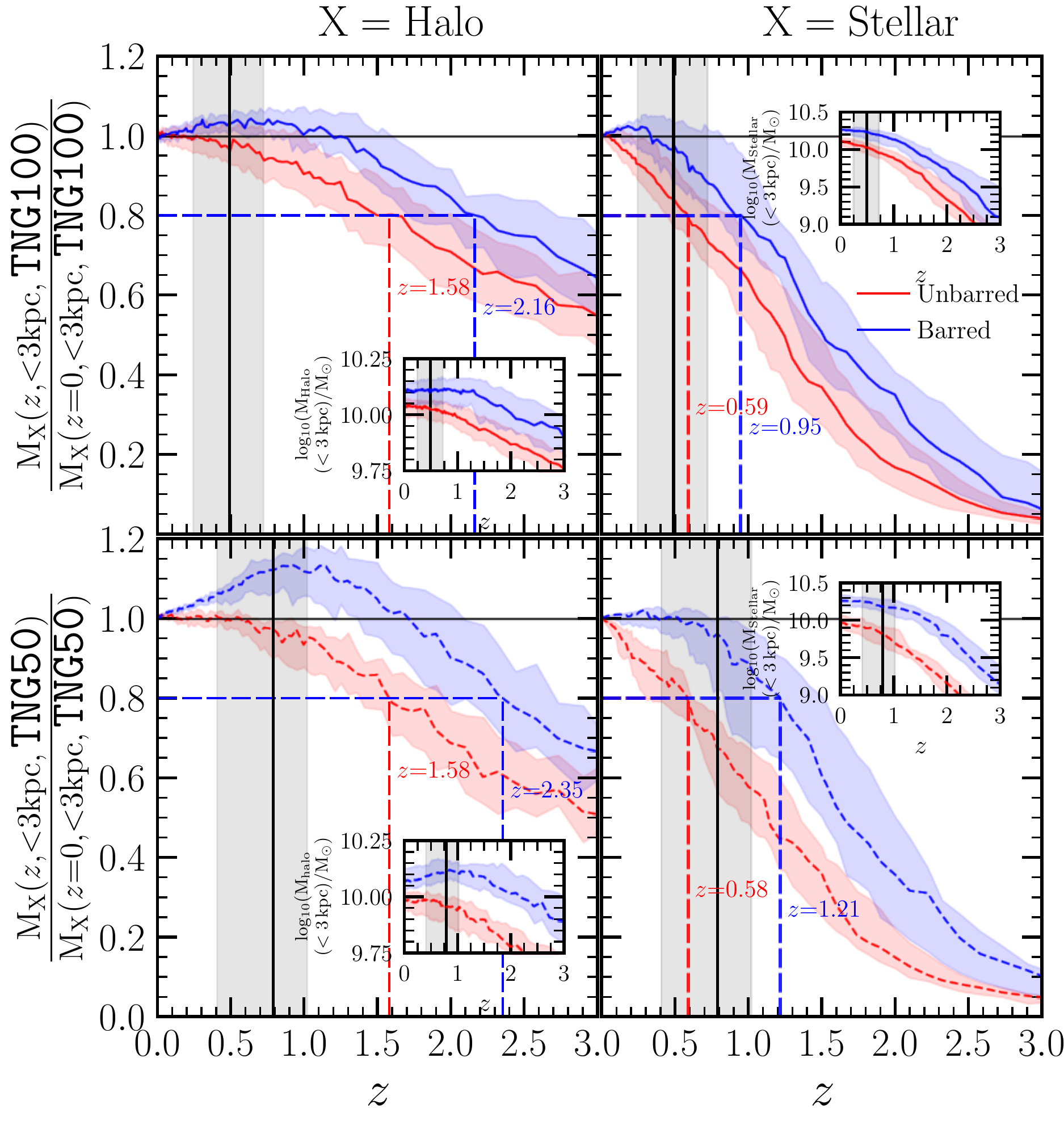}
\caption[]{Assembly of the stellar and halo component of barred (blue) and unbarred galaxies (red) within a physical distance of $\rm 3 \, kpc$. The inner plots display the total stellar and halo mass within a physical distance of $\rm 3 \, kpc$. In all the panels shaded areas represent the $\rm 32^{th} {-} \rm 68^{th}$ percentile. The vertical black line represents the median formation time of the bar structure of the whole barred galaxy sample. The grey area is the $\rm  32^{th}  \,{-}\, 68^{th}$ percentile of the bar formation distribution. Whereas the upper panels display the results for \texttt{TNG100}, the lower ones show the same for \texttt{TNG50}. In all the panels the horizontal lines mark the instants at which the DM halo/stellar component reached 80\% of their final values.}
\label{fig:Assembly_All_Bars_Unbars}
\end{figure}

Another important property in the study of disc structure is the vertical scale length, $ z_{\rm d}$, which gives information about how stars are distributed perpendicularly to the disc \citep[see e.g][]{Yoachim2006,Comeron2011}. To determine the vertical length, in the reference system aligned with the z-direction, we separate stellar disc particles from the bulge ones by computing the particle circularity parameter, $\eta$ \citep[][]{Abadi2003,Scannapieco2009,Marinacci2014}:
\begin{equation}\label{eq:Circularity}
\eta\,{=}\, \frac{J_{\rm z, \star}}{r v_c(r)}
\end{equation}
where $r$ is the star radial distance, $J_{\rm z,\star}$ the z-component of the angular momentum and $v_c(r)$ its circular velocity. Following \cite{Genel2015}, we define disc particles as the ones with $\eta\,{>}\,0.7$. The rest are tagged as bulge-like (\textit{hot component}). We highlight that further refinements of this disc-bulge kinematic distinction could be done (see, \citealt{Scannapieco2009,Marinacci2014,Du2019}, Zana et al. submitted), however, this level of refinement goes beyond the scope of this paper. Finally, the disc component is placed edge-on and we fit its surface density as:
\begin{equation}
    \rm \Sigma_d(z) \,{=}\, \frac{\Sigma_{\mathit{z}_0}^{d}}{2\mathit{z}_d} sech^2\left(\frac{\mathit{z}}{\mathit{z}_d}\right)
\end{equation}
where $\rm \Sigma_{\mathit{z}_0}^{d}$ and $\rm \mathit{z}_d$ are the central surface density and vertical scale length of the disc.

\begin{figure*}
\centering
\includegraphics[width=1\columnwidth]{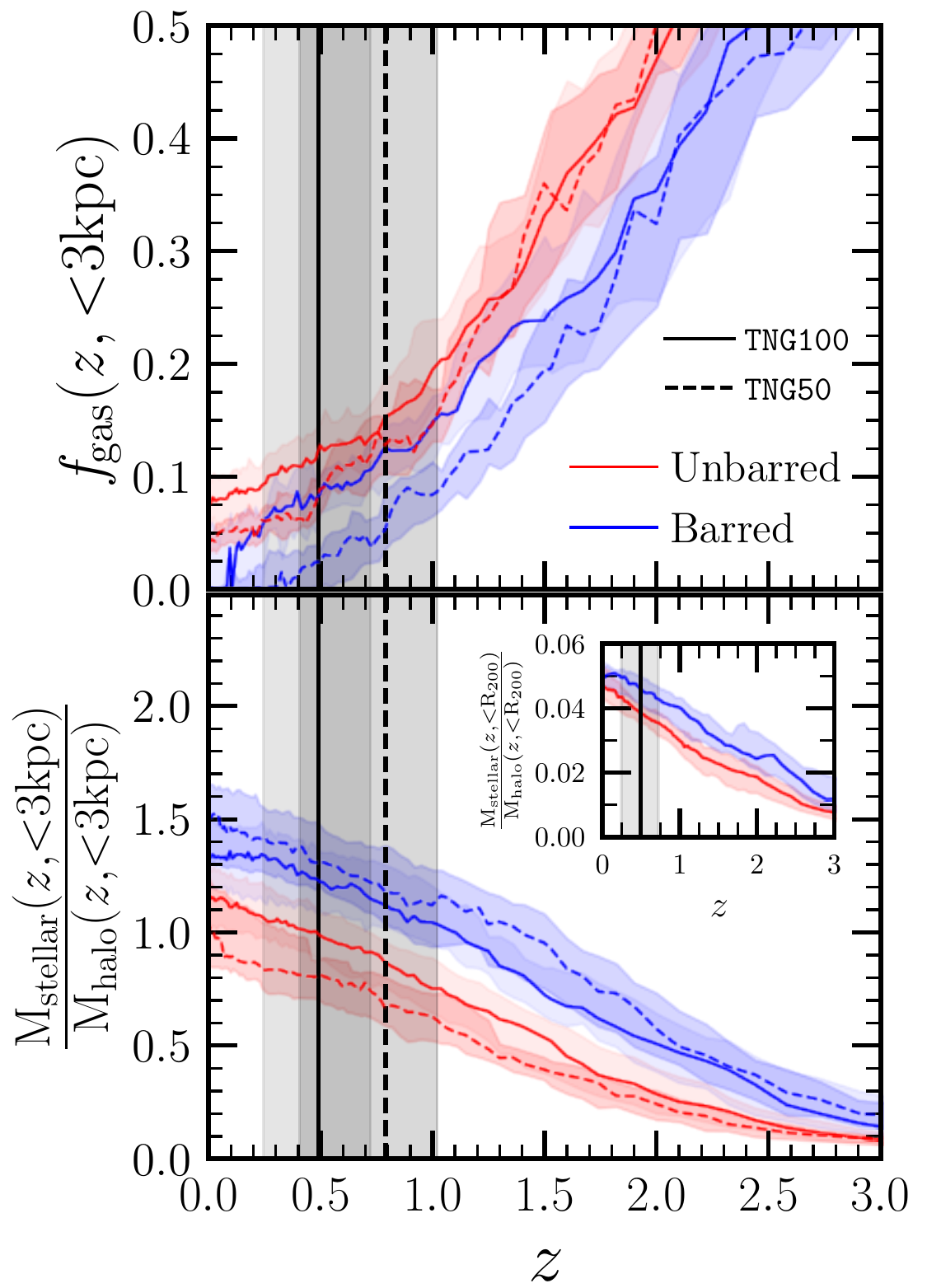}
\includegraphics[width=1\columnwidth]{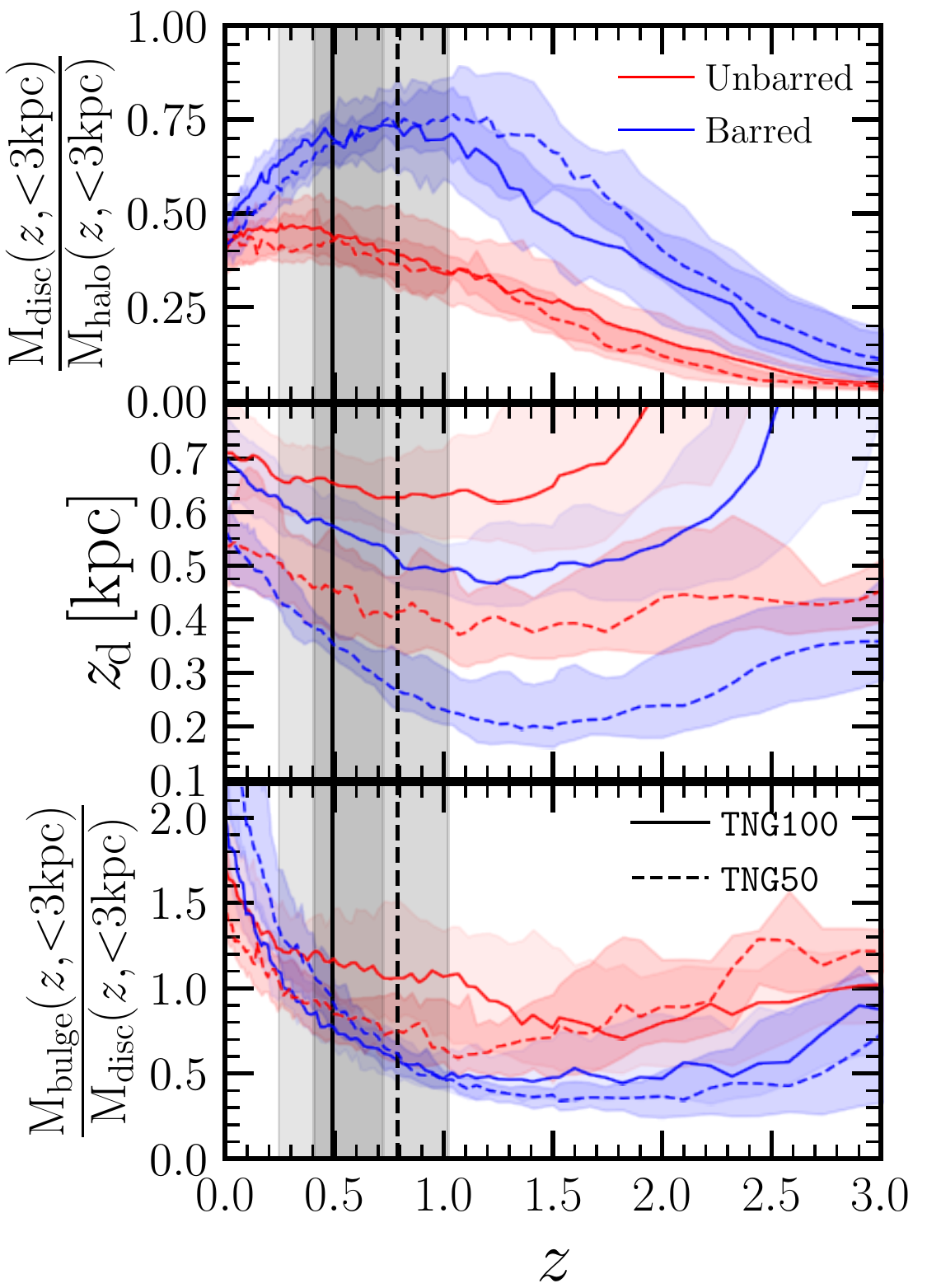}
\caption[]{\textbf{Left panels}: Gas fraction, $f_{\rm gas}$, and stellar-to-halo ratio at $\rm {<}\,3\, kpc$. The inset displays the median value of the ratio between stellar and DM mass within the dark matter virial radius, $\rm R_{200}$. \textbf{Right panels}: Disc-to-halo and bulge-to-disc ratio within $\rm {<}\,3\, kpc$ and the vertical scale length of the disc ($z_{\rm d}$). In all the plots, blue and red lines represent the median results for barred and unbarred galaxies whereas shaded areas are the $\rm 32^{th} {-} \rm 68^{th}$ percentile of the distribution. Solid and dashed lines represent the predictions for \texttt{TNG100} and \texttt{TNG50}. The vertical solid and dashed black lines represent, respectively, the median formation time of the bar structure of the whole \texttt{TNG100} and \texttt{TNG50} barred galaxy samples. Grey area shows the $\rm 32^{th} {-} \rm 68^{th}$ percentile of the bar formation distribution.}
\label{fig:Stellar_Halo_Ratio}
\end{figure*}

\section{The general properties of barred and unbarred galaxies} \label{sec:AssemblyBarsUnbars}

We start by analyzing the general properties of the barred and unbarred populations of galaxies in \texttt{TNG100} and \texttt{TNG50}. In Fig.~\ref{fig:Mdisc_Rdisc_Vmax_Bars} we present the evolution of the three quantities relevant in the \ec:  the stellar mass of the disc, $\rm M_{disc}$, the disc scale length, $R_{\rm d}$ and the maximum circular velocity of the system\footnote{The rotation curve of the system is computed considering all the components, namely stellar, dark matter and gas.}, $\rm V_{max}$ (see Eq.\ref{eq:Bar_formation_Efstathiou1982}). By construction, barred and unbarred galaxies have similar disc masses at $z\,{=}\,0$ ($\rm M_{disc}\,{\sim}\,10^{10.5}\, M_{\odot}$). However, barred galaxies assembled earlier and, at higher redshifts they display more massive discs. For instance, at $z\,{\sim}\,1$ the discs of barred galaxies are ${\sim}\,0.5\,\rm dex$ more massive than of unbarred galaxies, with similar values obtained in both the \texttt{TNG100} and \texttt{TNG50}.  Regarding $R_{\rm d}$, we can see the barred sample displays smaller values than the unbarred one: the discs of barred galaxies are more compact, likely because of their earlier formation time and their smaller DM halo spin, as we will discuss below and in Section~\ref{sec:ComparingProperties_BUR_UBUR_UBSR_UBUR}. While being consistent in the overall behaviour,  in this case we see some differences between \texttt{TNG100} and \texttt{TNG50}: in the \texttt{TNG100} the barred and unbarred samples have similar values of  $R_{\rm d}$ at high redshift ($z\,{\gtrsim}\,1$) and start diverging at lower times. We emphasize that $R_{\rm d}$ of both samples starts to differ at the bar formation time. Therefore, it is not clear whether this is a cause or a consequence of the bar formation. On the other hand, for the case of \texttt{TNG50}, barred and unbarred galaxies always show a constant offset of $\rm {\sim}\,1\,kpc$. Finally, differences are also seen in the maximum circular velocity. Regardless  of redshift, barred galaxies are hosted in systems with ${\sim}20\%$ larger $\rm V_{max}$ in the  \texttt{TNG100}, while differences can be as large as ${\sim}40\%$ for the \texttt{TNG50}.\\

Before investigating how these values translate in to the ability of the \ec{} to capture the stability of the disc, we further investigate the general properties of the two populations. 
In Fig.~\ref{fig:Assembly_All_Bars_Unbars} we present the assembly of the inner region ($\rm {<}\,3 \, kpc$) of the halo and stellar components. We highlight that the same trends are seen at $\rm 2\,kpc\,{<}\,\mathit{r}\,{<}\,5\,kpc$\footnote{To explore possible dependencies with the halo and stellar mass, we have compared the build-up of a sample of bars and unbarred galaxies matched with the same DM halo mass and stellar mass at $z\,{=}\,0$. The comparison showed that, regardless of mass, barred galaxies assembled their stellar and DM components earlier.}. Regarding the halo component (left panels), at $z\,{=}\,0$ the two samples display similar  final masses, with differences smaller than $\rm 0.1\, dex$ (see the inset). However, the redshift evolution is considerably different: the inner region of the halos of barred galaxies is significantly more massive at all redshifts. These differences in mass are reflected in the halo assembly, being faster for the barred sample. For instance, while the halos hosting barred galaxies aggregated $80\%$ of their mass within $\,3\, \rm kpc$ by $z\,{\sim}\,2$, those of unbarred galaxies did it by $z\,{\sim}\,1.6$. Interestingly, the central halo component  of both barred samples decreases at $z\,{<}\,1.5$, with this effect being more pronounced in the \texttt{TNG50} sample. Such depletion of dark matter in the central region can be due to the assembly of the stellar disc component. Indeed, \cite{Yurin2015} reported that the presence of discs can reduce the DM halo in the inner parts of the galaxy by a factor of 2. The authors attributed this depletion to the gravitational shocks that DM particles experience as they pass through the disc. Interestingly, this decrease takes place around the bar formation time, indicating that bar structures  might be also contributing to a redistribution of the dark matter component. Similar results were found by \cite{Algorry2017} when exploring the evolution of barred galaxies in the \texttt{EAGLE} simulation. Besides, these authors found that the DM redistribution correlates with the strength of the bar, being larger for the strongest ones. To check if this effect can explain the larger drop seen in the \texttt{TNG50}, in Fig.~\ref{fig:A2_tlookbacktime} we compared the values of $\rm A_2^{max}$ in \texttt{TNG100} and \texttt{TNG50}. Interestingly, we find that the median value of \texttt{TNG50} is $\rm A^{max}_2 \,{\sim}\,0.4$ whereas in \texttt{TNG100} it is $\rm A^{max}_2 \,{\sim}\,0.35$. While we can not exclude the fact that resolution effects can  affect the $\rm A^{max}_2$ values \citep[see e.g][]{Frankel2022}, these results support the correlation between the bar strength and the efficacy of dark matter depletion seen in \cite{Algorry2017}. \\ 

The stellar component follows a similar behavior to that of dark matter: barred galaxies have a larger mass content and faster assembly in the inner region with respect to unbarred galaxies. The early build up of galaxies that eventually develop a bar component was also reported by \cite{RosasGuevara2019}, who related it with the fact that barred galaxies experienced earlier and more efficient star formation episodes than unbarred ones. Indeed, when we look at the gas fraction\footnote{We define the galaxy gas fraction as $f_{\rm gas}({<}\,r) \,{=}\, \rm M_{gas}({<}\,\mathit{r}) / (M_{gas}({<}\,\mathit{r}) + M_{stellar}({<}\,\mathit{r}))$} (top panel of Fig.~\ref{fig:Stellar_Halo_Ratio}) we see that, regardless of redshift and radius, the barred sample exhausts the nuclear gas reservoir earlier than the unbarred one. Indeed, the recent observational study of \cite{FraserMcKelvie2020} supports this scenario. The authors showed that, regardless of mass, star formation histories of barred galaxies peak at earlier times than their unbarred counterparts, suggesting that the bar hosts build up their stellar component at higher redshifts. Interestingly, when comparing the \texttt{TNG100} and \texttt{TNG50} barred galaxies, we can see that the latter consumed their gas reservoir faster. This could explain why bar structures were born earlier in \texttt{TNG50} ($z\,{\sim}\,0.7$) than in \texttt{TNG100} ($z\,{\sim}\,0.5$, see Fig.~\ref{fig:A2_tlookbacktime}): faster gas consumption can lead to a faster disc assembly which, in turn, can induce bar instability leading to an earlier formation of a bar structure. On top of this, the faster gas consumption seen in \texttt{TNG50} could shed light on why the bar structures formed in that simulation are stronger than their counterparts in \texttt{TNG100}. By performing several simulations with different gas fractions, \cite{Athanassoula2013} reported that bars in gas-poor galaxies were longer and reached a higher strength than the ones hosted in gas-rich galaxies.\\

To explore the relative assembly between the dark matter halo and the stellar component, in the lower panel of the left column of Fig.~\ref{fig:Stellar_Halo_Ratio} we present the redshift evolution of the ratio between the stellar and the halo mass (hereafter stellar-to-halo) for bars and unbarred galaxies at $\rm {<}\, 3\,kpc$. As we can see, regardless of redshift, the stellar-to-halo content is always larger for the barred sample (typically a factor ${>}\,1.2$) in both simulations. The difference is already present before bar formation and it persists down to $z=0$. Such  differences are still seen when we  examine the total stellar-to-halo ratio within the halo virial radius (inset plot of Fig.~\ref{fig:Stellar_Halo_Ratio}). Similar trends were already reported by observational and theoretical works. On the observational side, \cite{Cervantes2017}, by analyzing galaxies in the SDSS-DR7, found an increasing trend of the bar fraction with larger stellar-to-halo mass ratios \citep[see also][]{Cervantes2015}. Similar results were shown by \cite{DiazGarcia2016} using the $\rm S^4G$ survey \citep{Sheth2010}. On the theoretical side, \cite{Valenzuela2003} studied the transfer of angular momentum between halos and bars using N-body simulations of isolated disc galaxies. Interestingly, the authors found that the transfer of angular momentum from the bar to the outer regions of the disc leads to an increase (decrease) of the stellar (halo) mass content in the centre of the galaxy. On the same line we find the work of \cite{Fragkoudi2020} in which the authors, by analyzing the the zoom-in \texttt{AURIGA} simulations, found that the hosts of barred galaxies display higher stellar-to-dark matter ratios than what is expected from the abundance matching relation. Finally, similar trends were found by \cite{RosasGuevara2021} who reported that at fix stellar (or DM) mass, \texttt{TNG50} barred galaxies were systematically more stellar dominated than the unbarred population.\\

Finally, we take a closer look at the evolution of disc and bulge components\footnote{The division between disc and bulge-like particles has been done as explained in Section~\ref{sec:VerticalScaleLength}} in the right panel of Fig.~\ref{fig:Stellar_Halo_Ratio}. We first show the evolution of the disc-to-halo ratio in the central region ($\rm {<}\, 3\,kpc$), seeing that barred galaxies display systematically larger values than unbarred ones, except for $z=0$. Before bar formation, the disc-to-halo ratio of barred galaxies rises steeply. These results support the results of \cite{Mayer2004} and \cite{DeBuhr2012}, who found through numerical simulations of isolated galaxies that the disc-to-halo mass ratios within the typical disc radius can be a fundamental factor for determining the development of barred structures in disc-dominated galaxies.  However, after bar formation  ($z\,{<}\,0.5$ for \texttt{TNG100} and $z\,{<}\,0.7$ for \texttt{TNG50}), the disc-to-halo ratio of barred galaxies drops. This effect can be caused by a combination of factors. One could be that the bar structure triggers a buckling of the stellar orbits, leading to the formation of a pseudobulge component \citep{Pfenniger1990,Bureau1999,Combes2009,Kormendy2013,Fragkoudi2017}. Another alternative could be that the bar structure itself is dominated by more radial orbits, which are likely to appear in a mildly rotating component rather than in a cold disc (i.e, ${\eta}\,{>}\,0.7$). This would cause our morphological decomposition to classify some of the stellar particles of the bar as a bulge (\textit{hot}) component. Besides the different behaviour in the disc-to-halo ratio, the discs of barred and unbarred galaxies are distributed differently also in the vertical axis ($z_{\rm d}$). As shown, barred galaxies display colder discs than unbarred ones, with vertical heights up to $1.5$ times smaller. We stress that \texttt{TNG50} displays smaller values of $z_{\rm d}$ than \texttt{TNG100}, principally as a result of resolution effects. As shown by \cite{Pillepich2019} (see their Figure B.2), lower resolution simulations have puffier stellar discs. Despite the differences between \texttt{TNG100} and \texttt{TNG50}, our results agree with the theoretical work of \cite{Athanassoula1986} and \cite{Athanassoula2003} which reported that dynamically hot discs can delay or even suppress the formation of bars as a consequence of large random motions hindering the growth of bar modes. Observational evidence of such an effect can be found in \cite{Sheth2012}: analyzing massive galaxies at $z\,{<}\,0.8$ the authors found that bar structures are preferentially in galaxies with massive and and dynamically cold discs.\\

Lastly, concerning the evolution of the bulge-to-disc ratio, we see that at $z\,{\gtrsim}\,0.5\,{-}\,0.7$ unbarred galaxies display values larger than $0.5$, pointing out that the bulge component dominated the inner parts of the galaxy. In contrast, barred galaxies have bulge-to-disc ratios smaller than $0.5$, with the disc governing the inner galaxy. At $z\,{\lesssim}\,0.5\,{-}\,0.7$, whereas no dramatic change is seen in the unbarred sample, barred galaxies underwent a change of trend. Specifically, the bulge gains more relevance as bar formation triggers the development of a bulge-like component (i.e, a structure with a \textit{hotter} kinematics than the disc). Finally, we highlight that the decreasing trend seen in both samples at $z\,{\gtrsim}\,1.5$ is likely caused by the fact that galaxies start with an irregular morphology, causing our morphology distinction to classify most of the stellar particles as bulge-like (or hot component). Later on, as the galaxy evolves, it develops a well-defined disc component and creates a bulge structure through secular and merger processes. We refer the reader to Zana et al. (submitted) for further details about the assembly of discs and bulges in \texttt{TNG} simulations.\\

Before moving on to the direct test of the \ec{} for detecting the capability of a galaxy to form a bar, we conclude from this first analysis that the bar properties and their host galaxies in the \texttt{TNG50} and \texttt{TNG100} simulations generally confirm several trends derived in earlier works: galaxies that eventually develop a bar assemble at  earlier times, are characterized by larger stellar-to-halo ratios in the inner regions and their discs are more massive, but smaller and colder, with respect to unbarred galaxies. Early global assembly and a fast build up and compactness of the disc seem to be key to bar formation.  Moreover, we also find indications that the bar is able to redistribute the dark matter component in the inner parts of the galaxy, in particular diminishing its concentration.

\section{Accuracy of the ELN-criterion} \label{sec:ECriterionPerformance}

\begin{figure}
\centering
    \includegraphics[width=1\columnwidth]{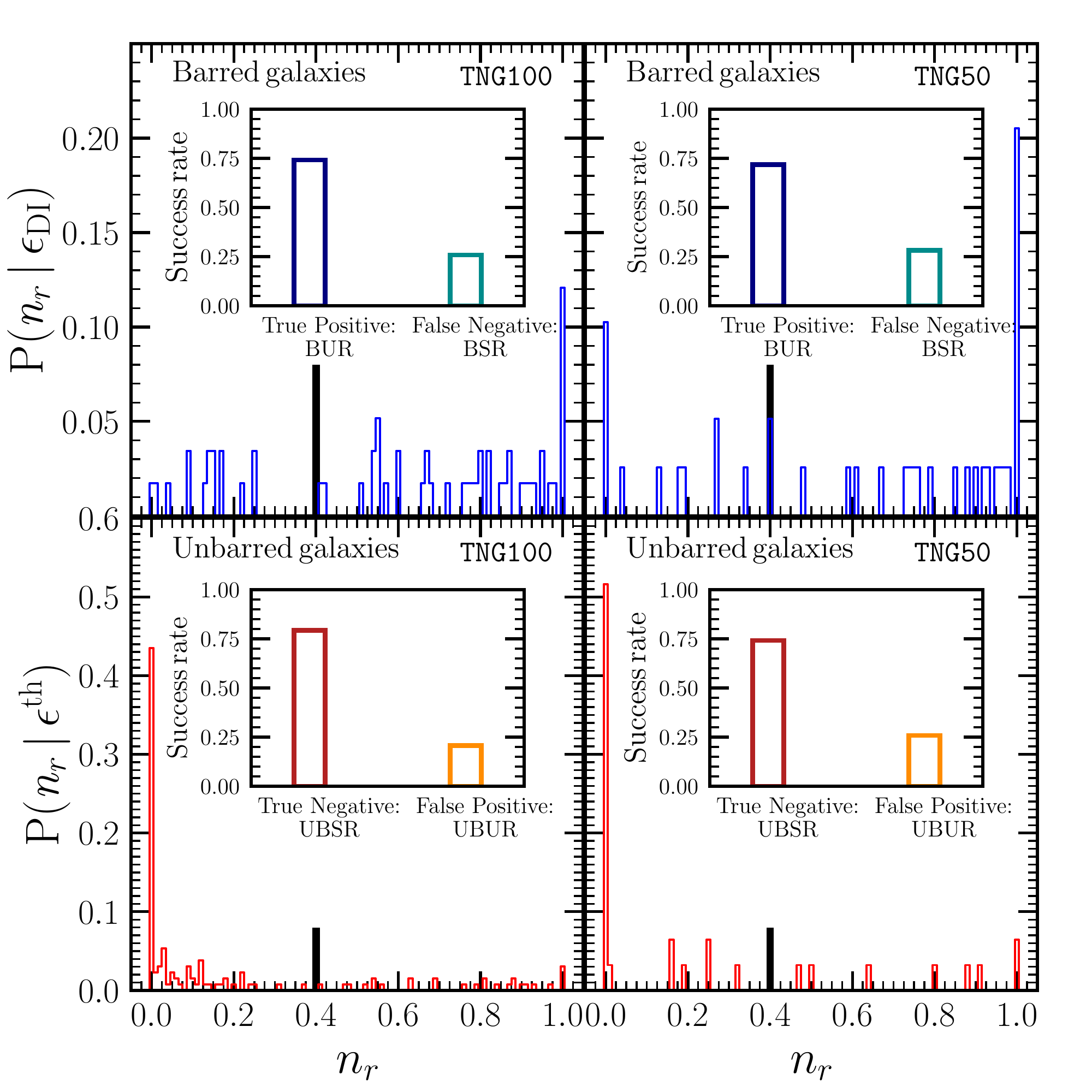}
\caption{Distribution of $n_r$ for all the barred (top) and unbarred (bottom) galaxies. Left and right panels represent the predictions for \texttt{TNG100} and \texttt{TNG50}, respectively. The solid black vertical line highlights the threshold $n_r\,{=}\,0.4$, which we use to determine the success rate of the \ec{}, assuming a threshold of $\epsilon=1.1$. The derived success rate is shown in the inset. True positives, i.e barred galaxies correctly identified by the criterion, are labelled BUR (bar in the unstable region). On the other hand, false  negatives, i.e barred galaxies in which the criterion does not find bar instability, are labelled as BSR (bars in the stable region).  Similarly, unbarred galaxies properly identified by the criterion (true negative) are called UBSR (unbarred in the the stable region) and unbarred galaxies wrongly considered unstable are called UBUR (unbarred in the unstable region).}
\label{fig:Success_rate}
\end{figure}

After exploring the global differences between bars and unbarred galaxies and determining the evolution of the quantities employed by the \ec, we now examine its ability to determine whether a bar actually forms. To do this, we do not evaluate the criterion only at the snapshot corresponding to bar formation, but over a broader time interval. We define a new variable, $n_r$, which measures the fraction of time spent by the galaxy in the bar unstable region. Specifically, it is defined as:
\begin{equation} \label{eq:nr}
n_r =  \frac{1}{t_f^{\rm bar}} \sum_{i {=} \rm snp(\mathit{z}_{\mathit{f}}^{bar})}^{\rm snp(\mathit{z}\,{=}\,0)} \Delta t (i\,,\,i+1|\epsilon\,{\leq}\, \epsilon^{\rm th}) ,
\end{equation}
where $\Delta t (i\,,\,i+1|\epsilon\,{\leq}\, \epsilon^{\rm th})$ corresponds to the time interval between the snapshot $i$ and subsequent one, $i+1$, in which the galaxy $\epsilon$ value is smaller than $\epsilon^{\rm th}=1.1$ (see Eq.~\ref{eq:Bar_formation_Efstathiou1982}). The index $i$ runs from the snapshot of the bar formation until the last simulation snapshot (i.e $z\,{=}\,0$). Therefore, $n_r \, {=} \, 0$ for galaxies that never satisfy the condition of $\epsilon\,{\leq}\,1.1$ at a snapshot after bar formation ($\mathrm{snp \,{\geq} \, snp_{\mathit{f}}^{bar}}$). On the contrary, $n_r \, {=} \, 1$ for galaxies that always satisfy the instability criterion. In the top panels of Fig.~\ref{fig:Success_rate} we show the distribution of $n_r$ for barred galaxies in \texttt{TNG100} (left) and \texttt{TNG50} (right). Most barred galaxies have a large value of $n_r$, with a clear maximum at $n_r\,{\sim}\,1$ (more pronounced in \texttt{TNG50}), showing that  the majority of barred galaxies satisfy the \ec{} for a significant fraction of time after bar formation. However, there is a small number of  barred galaxies with low values of  $n_r$, pointing out that some barred galaxies do not satisfy $\epsilon{\leq}1.1$ for most of the snapshots after bar formation. On the other hand,  in the lower panel of Fig.~\ref{fig:Success_rate} we present the  $n_r$ distribution for unbarred galaxies. For this sample, we have decided to use as $\rm snp(\mathit{z_f}^{bar})$ the median $\rm snp(\mathit{z_f}^{bar})$ of the barred galaxies (see discussion of Section~\ref{sec:Bars_NoBars_and_Fit}). The vast majority of unbarred galaxies are characterized by low values of  $n_r$, indicating that the \ec{} is generally able to capture the stability of disc galaxies. \\

Based on the distributions shown in Fig.~\ref{fig:Success_rate} we have chosen a threshold of $n_r$ ($n_r^{\rm th}$) equal to $0.4$ to define the cases we consider that the \ec{} successfully identifies disc stability. The value for this threshold is arbitrary, and we have checked that the results do not significantly change when varying $n_r^{\rm th}$ between $0.3\,{-}\,0.5$. In the inset plots of Fig.~\ref{fig:Success_rate} we present the derived success rate of the \ec{}. For barred galaxies, we obtain a success rate of ${\sim}\,74$/$72\%$ with a failure fraction of ${\sim}\,26$/$28\%$ for the \texttt{TNG100}/\texttt{TNG50}, respectively. 
From hereafter, galaxies in the former case are going to be called \textit{bars in the unstable region} (BUR, true positive) while the ones in the latter are going to be tagged as \textit{bars in the stable region} (BSR, false negative). Regarding the unbarred sample, the number of galaxies correctly identified as unbarred by the \ec{} are ${\sim}\,79$/$75\%$, while ${\sim}\,21$/$25\%$ are misclassified as being unstable, for \texttt{TNG100} and \texttt{TNG50}, respectively. In analogy with the barred sample, from now on we refer to the unbarred galaxies correctly classified as \textit{unbarred in the stable region} (UBSR, true negative) while the ones misclassified are going to be tagged as \textit{unbarred in the unstable region} (UBUR, false positive).\\

To explore the robustness of the choice of $1.1$ as threshold value in the \ec{}, we also calculated the success rate varying the adopted threshold. As expected, the number of correctly identified bars increases when adopting a larger threshold, at the expense of a larger number of unbarred galaxies being  tagged as unstable.  For example, if we adopt $\epsilon^{\rm th}\,{=}\,1.2$, a value also used in the literature \citep{Lagos2018,IzquierdoVillalba2019}, the number of barred galaxies correctly identified  increases up to ${\sim}\,85/77\%$ for \texttt{TNG100}/\texttt{TNG50}, while the number of correctly identified unbarred galaxies  drops to $\,{\sim}\,67/55\%$ for \texttt{TNG100}/\texttt{TNG50}. 
We find that the best balance between correctly identified barred and unbarred galaxies is reached when adopting the threshold of $1.1$ as originally proposed by \cite{Efstathio1982} \citep[see also][]{Yurin2015}. 
We have also explored how the success rate of the \ec{} changes depending on the adopted time of bar formation i.e, $\rm snp(\mathit{z_f}^{bar})$. We have found that moving to any other snapshot within $t_{f}^{\rm bar} \, {\pm} \, 40 t_{\rm dyn}$ (where $t_{\rm dyn}$ the disc dynamical time at bar formation time), the success rate for barred and unbarred galaxies does not change. 
Finally, we have explored the success rate when the summation of  Eq.~\ref{eq:nr} is performed only in a few snapshots around the bar formation, specifically over the snapshots within $10\,t_{\rm dyn}$ before and $40\,t_{\rm dyn}$ after the bar formation time. Again, we find only minor differences with the success rates obtained above.\\

\begin{figure}
\centering
\includegraphics[width=1\columnwidth]{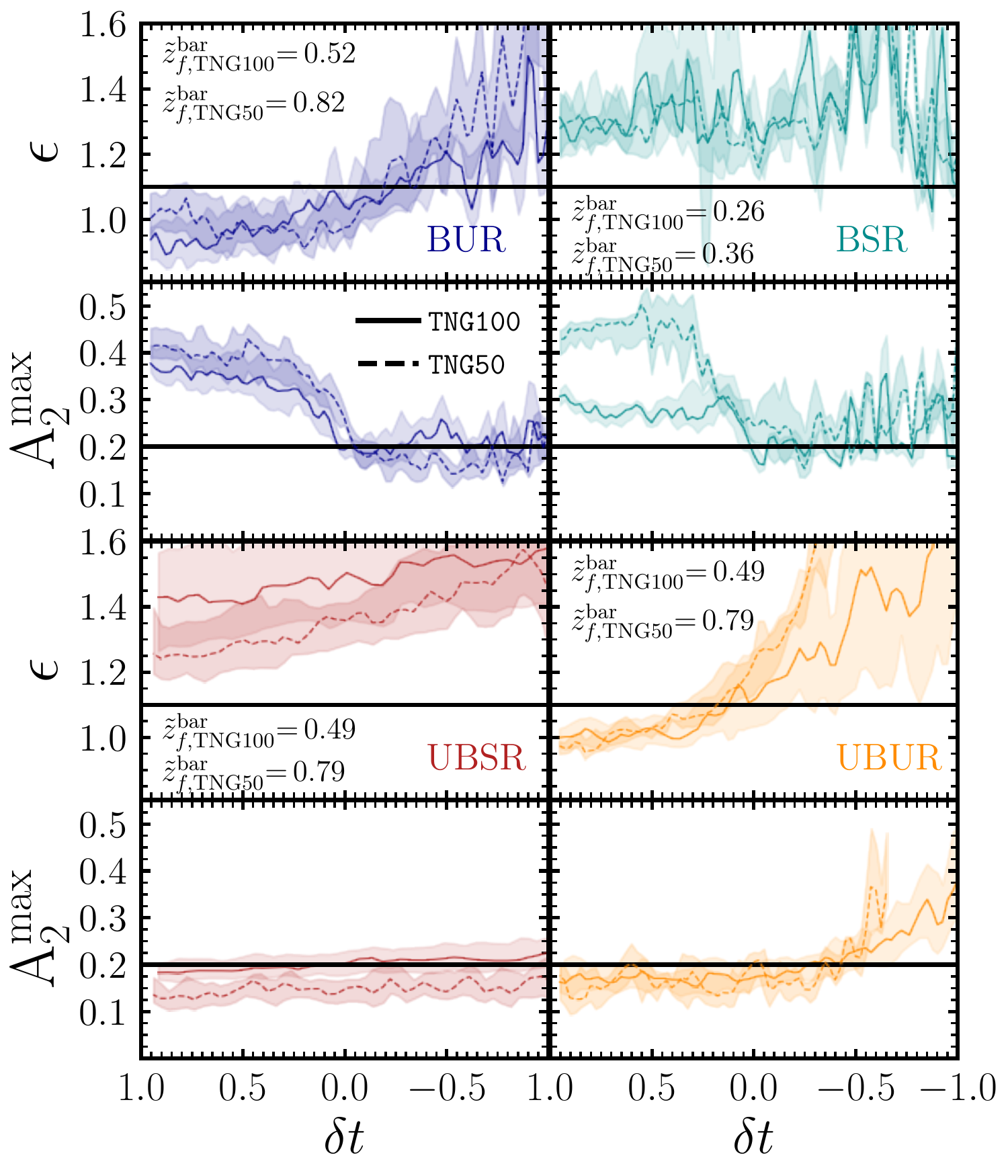}
\caption[]{\textbf{Upper panels}: Dark blue thick lines represent the evolution of the median values of $\epsilon$  and $\rm A_2^{max}$  for the \textit{barred in the unstable region sample} (BUR - true positives). Cyan lines represent the same but for the \textit{barred in the stable region} (BSR - false negatives). Shaded areas display the percentile $\rm 32^{th}{-}68^{th}$.  In each panel we  show the $\Bar{z}_f^{\rm bar}$, i.e., the median formation redshift for each sample.  \textbf{Lower panels}: The same as the upper panels but for the \textit{unbarred in the stable region sample} (red, UBSR - true negative) and \textit{unbarred in the unstable region sample} (orange, UBUR - false negative). In all the plots, solid and dashed lines represent the results for \texttt{TNG100} and \texttt{TNG50}, respectively.  For unbarred galaxies, times are normalized using the median formation time of the whole barred sample of the corresponding simulation.}
\label{fig:A2_epsilon_BUR_BSR}
\end{figure}

In the upper panels of Fig.~\ref{fig:A2_epsilon_BUR_BSR} we present, for both \texttt{TNG100} and \texttt{TNG50}, the median evolution of $\rm A_2^{max}$ and $\epsilon$ as a function of $\delta t$ (see Eq.~\ref{eq:Normalizet_time_bar}) for the BUR and BSR samples (true positive and false negative, respectively). As we can see, the first sample displays a good correlation between  $\epsilon$ and the bar strength evolution, i.e.,  $\epsilon$ decreases down to values smaller than $1.1$ when  $\rm A_2^{max}$ increases above the threshold value of $0.2$ after the time of bar formation ($\delta t\,{\sim}\,0$). On the other hand, the galaxies belonging to the BSR sample are characterized by $\epsilon$ values systematically larger than $1.1$ before and after bar formation.  We note here that the BUR sample has a median bar  formation time ($z\,{\sim}\,0.5/0.8$ for \texttt{TNG100}/ \texttt{TNG50}) larger than the one of the BSR sample ($z\,{\sim}\,0.25/0.35$ for \texttt{TNG100}/ \texttt{TNG50}). On top of this, the evolution of the $\rm A_2^{max}$ median value is different for the BSR and the BUR galaxies. While \texttt{TNG100} predicts that the BUR sample displays systematically larger  $\rm A_2^{max}$ values than BSR, \texttt{TNG50} predicts the opposite. These differences could hint at the fact that the barred galaxies of the BSR sample might have a different evolution than the one of the BSR sample and/or their bars form via different processes. In the bottom panels of the same figure, we show the median evolution of $\rm A_2^{max}$ and $\epsilon$ as a function of $\delta t$ for the unbarred sample divided into UBSR (true negative) and UBUR (false positive) samples. As mentioned before, for unbarred galaxies $\delta t$ has been computed  using the median formation time of all barred galaxies.
 As we can see, the UBSR galaxies are characterized by values of $\epsilon$ above the threshold value of the \ec{} and a flat $\rm A_2^{max}$ evolution with time, confirming the stability of the sample. For the UBUR sample, instead, we find a similar evolution for  $\rm A_2^{max}$, but  $\epsilon$  decreases with time, reaching values below the disc stability threshold. We note that in some cases, in particular at early times,  $\rm A_2^{max}$ can be larger than 0.2. Those fluctuations in   $\rm A_2^{max}$  are not related to a bar structure given that the phase $\varphi$ associated with that $\rm A_2^{max}$ within the bar length is not constant. Instead, large values of $\rm A_2^{max}$ can be caused by any non-axisymmetric structure such as spiral arms or interactions with small satellites. \\

Based on the analysis performed in this section, we can conclude that, despite its simplicity, the \ec{} is able to correctly identify the formation or absence of a barred structure in the majority of the cases. This in agreement with the results of \cite{Yurin2015} who found a correlation between the outcome of the \ec{} and the $z\,{=}\,0$ bar strength by analyzing simulations of Milky-Way type galaxies. However, the recent work by \cite{Algorry2017} (\texttt{EAGLE} simulation) and \cite{Marioni2022} (zoom-in simulations) pointed towards the opposite direction, suggesting that \ec{} is an incomplete indicator of disc instability. The main difference between our analysis and the one performed in these works consists in that they check the \ec{} just at the bar formation time, defined as the moment in which the galaxy fulfils $\rm A_2^{max}\,{>}\,0.2$. Probably, averaging the \ec{} over several snapshots (as we do) might also improve the robustness of the criterion in their simulations. Instead of being an instantaneous process, the development of a bar structure can take several galaxy dynamical times. On the observational side, the recent paper of \cite{Romeo2022} explored the capability of \ec{} to detect the presence of bar structures in a sample of 91 disc dominated galaxies with $10^{6.5}\,{-}\,10^{11.5}\, \rm M_{\odot}$. The authors found that \ec{} was only successful in 55\% of the cases. Even though these results differ from our findings, a direct comparison between their work and our results is not straightforward given the different ranges of mass probed by our analyses and our pre-selection of strong barred galaxies. Trough private communication with the authors, we checked that in the same mass range probed by this work the results of \cite{Romeo2022} showed that the large majority of barred galaxies reside in the bar-unstable region, in agreement with our results. However, non-barred galaxies tend to be far from stability according to the \ec{}.\\

Despite the overall success, the stability/instability of almost one quarter of the galaxies is not correctly captured by the \ec{}. In what follows, we study the differences in the properties of the BUR and UBUR samples and the BSR and UBSR samples, with the goal of identifying the reasons why the \ec{} fails in certain instances. 


\subsection{The stellar and dark matter component of stable and unstable galaxies} \label{sec:ComparingProperties_BUR_UBUR_UBSR_UBUR}

\begin{figure}
\centering
\includegraphics[width=1.\columnwidth]{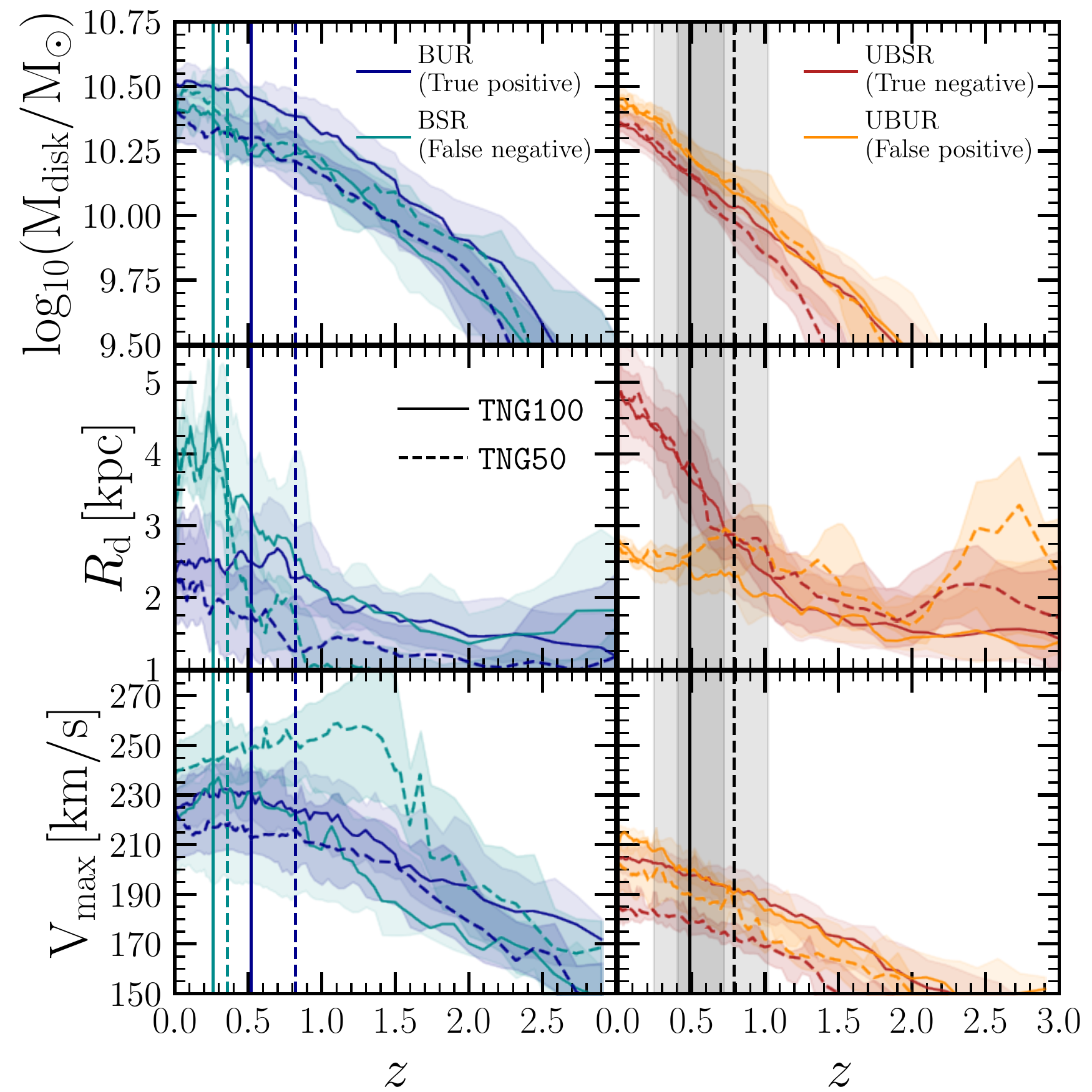}
\caption{Evolution of the stellar disc ($\rm M_{disc}$), stellar scale length radius ($R_{\rm d}$) and maximum circular velocity ($\rm V_{max}$) for the BUR (dark blue), BSR (cyan), UBUR (orange) and UBSR (red) sample. Solid and dashed lines represent the results for \texttt{TNG100} and \texttt{TNG50}, respectively. While lines display the median value, the shaded areas display the $\rm 32^{th}\,{-}\,68^{th}$ percentiles. For barred galaxies, the solid vertical lines highlight the median redshift of bar formation for each barred sample (we do not add the percentiles to avoid overcrowding). For unbarred galaxies, the solid vertical line highlights the median redshift of bar formation of all barred galaxies.}
\label{fig:Mdisc_Rdisc_Vmax_BUR_BSR_UBUR_UBSR}
\end{figure}

We start by looking at the redshift evolution of the physical quantities used in the \ec{}, $\rm M_{disc}$, $R_{\rm d}$ and $\rm V_{max}$, as in Fig.~\ref{fig:Mdisc_Rdisc_Vmax_Bars}, but now  separately for the  BUR/BSR (barred) and UBUR/UBSR (unbarred) populations. 
We show this in Fig.~\ref{fig:Mdisc_Rdisc_Vmax_BUR_BSR_UBUR_UBSR}. Regarding the barred sample (left column), in \texttt{TNG100} BUR galaxies (true positive) have generally more massive discs and larger $\rm V_{max}$ than the BSR ones (false negative). However, these trends disappear in \texttt{TNG50} where both samples display comparable $\rm M_{disc}$ and the BSR sample  reaches  large median $\rm V_{max}$. Despite these small differences, \texttt{TNG100} and \texttt{TNG50} agree that BUR galaxies have more compact discs, with differences of up to $\rm 1\, kpc$ with respect to BSR at $z_{f}^{\rm bar}$. The difference in the scale length of the disc is what primarily drives the different evolution of the $\epsilon$ parameter in the two samples. While $\rm V_{max}$ and  $\rm M_{disc}$ show relatively small variations between BUR/BSR and UBUR/UBSR (${\sim}\,10\,{-}\,20\%$), $R_{\rm d}$ can nearly double its value, which implies a change in $\epsilon$ by a factor up to $1.4$. 
Concerning the unbarred samples (right column of Fig.~\ref{fig:Mdisc_Rdisc_Vmax_BUR_BSR_UBUR_UBSR}), UBSR (true negative) and UBUR (false positive) present similar redshift evolution in $\rm M_{disc}$ and $\rm V_{max}$, but they show significant differences for $\rm \mathit{R}_d$ at $z\,{<}\,1$. While the scale lengths of UBUR galaxies experience a very mild evolution ($\rm \mathit{R}_d \, {\sim}\,2.5 \, kpc$) those of UBSR increase rapidly, reaching by $z\,{\sim}\,0$ values of $\rm \mathit{R}_d \, {\sim}\,4.5 \, kpc$. These trends  are found in both \texttt{TNG100} and \texttt{TNG50}. As for the barred sample, the difference  in the scale length of the disc is what causes the different evolution of the $\epsilon$ parameter for UBSR and UBUR. UBUR galaxies are very compact and have low values of $\epsilon$, thus, according to the \ec, they should be prone to instabilities. Comparing the BUR (true positive) and the UBUR (false positive) sample (i.e the two galaxy samples with the smallest radii), we see that the principal difference is seen for disc mass assembly, which happens earlier in the barred sample. We calculated the average formation time of the DM and stellar component of the two samples, and we found that the DM halos (stellar component) of BUR galaxies aggregated $80\%$ of their $z\,{=}\,0$ mass at ${<}\,3\,\rm kpc$ by $z\,{\sim}\,2$ ($z\,{\sim}\,1$) whereas UBUR ones did it ${\sim}\,1.5\rm \, Gyr$ later ($z\,{\sim}\,1.5$ and $z\,{\sim}\,0.5$ for the DM and stellar component, respectively). To summarize, Fig.~\ref{fig:Mdisc_Rdisc_Vmax_BUR_BSR_UBUR_UBSR} points out that $R_{\rm d}$ is the main property that leads to failures of the \ec. Large values of $R_{\rm d}$ are what cause some  barred galaxies to be misclassified as unbarred, and, conversely, the misclassified unbarred galaxies have untypically small values of the scale length.\\

\begin{figure}
\centering
\includegraphics[width=1.\columnwidth]{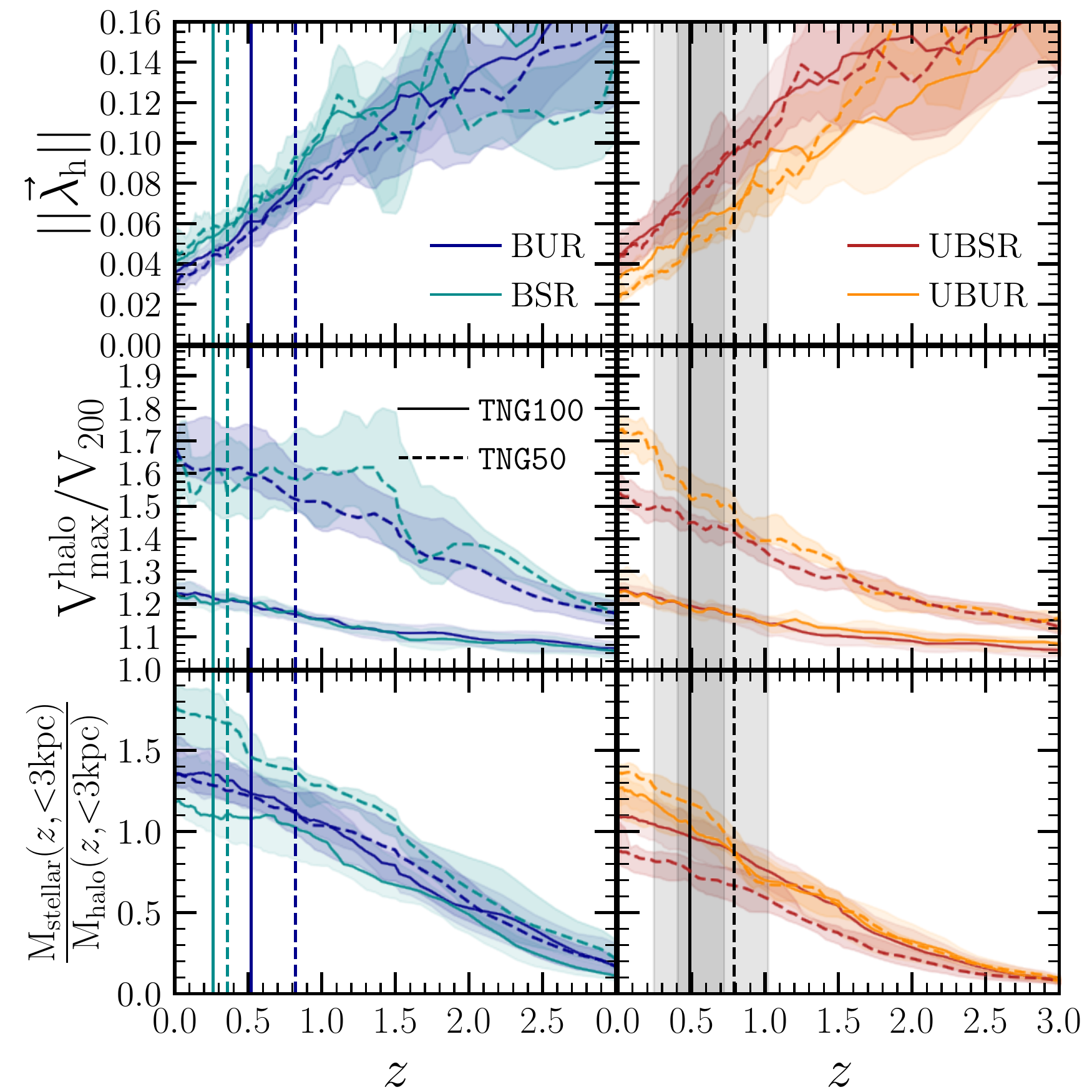}
\caption{Evolution of the DM halo spin modulus ($||\vec{\lambda}_{\rm h}||$), halo concentration ($\rm V_{max}^{\rm halo}/V_{200}$) and ratio between stellar and halo mass at $\rm {<}\,3\,kpc$. The color coding, line styles and vertical lines are the same as for Fig.~\ref{fig:Mdisc_Rdisc_Vmax_BUR_BSR_UBUR_UBSR}. }
\label{fig:Lh_Concentration_Mh_Ms_BUR_BSR_UBUR_UBSR}
\end{figure}

To explore the origin of the differences in $R_{\rm d} $ in the samples of barred and unbarred galaxies, in the right panel of Fig.~\ref{fig:Lh_Concentration_Mh_Ms_BUR_BSR_UBUR_UBSR} we look at the modulus of the DM halo spin, ${|}{|}\vec{\lambda}_{\rm h}{|}{|}$, computed following \cite{Bullock2001}:
\begin{equation}
\vec{\lambda}_{\rm h} \,{=}\,\frac{1}{\sqrt{2}}\frac{\sum_j \mathrm {M}_j \vec{r}_j\,{\times}\,\vec{\mathrm{v}}_j}{\rm M_{200}\,V_{200}\,R_{200}} 
\end{equation}
where $\mathrm{M}_{j}$, $\vec{r}_j$ and $\vec{\mathrm{v}}_j$ are the mass, radius and velocity of the $j{-}{\rm th}$ dark matter particle. $\rm M_{200}$, $\rm V_{200}$ and $\rm R_{200}$ correspond to the virial mass, velocity and radius of the subhalo, respectively. The samples with the largest $\rm \mathit{R}_d$, i.e UBSR and BSR (true and false positive, respectively), are the ones with the highest ${|}{|}\vec{\lambda}_{\rm h}{|}{|}$ parameter and no significant differences are seen between \texttt{TNG100} and \texttt{TNG50}. We highlight that these trends are important at low redshift ($z\,{<}\,1$) given that at high redshift the scatter dominates and the distributions of the four samples do not differ significantly. These results are in agreement with \cite{Mo1998} which reported that exponential discs embedded inside \textit{Navarro-Frenk-White} DM halos display scale lengths which scales proportionally to ${|}{|}\vec{\lambda}_{\rm h}{|}{|}$ (see Eq.10 of \citealt{Mo1998}). Therefore, the different $R_{\rm d}$ evolution between BUR/BSR and UBUR/UBSR seems to be the outcome of the different spinning DM halo in which these samples are hosted. Such correlation between galactic sizes and ${|}{|}\vec{\lambda}_{\rm h}{|}{|}$ has been debated during in the recent years. For instance, \cite{Jiang2019}, by using different zoom-in simulations of central and massive galaxies (larger or equal than Milky-Way type galaxies) found that the halo spin is not an useful predictor of the galaxy size, rather the halo virial radius is the property which displays the better correlation. This lack or correlation was also reported by \cite{Scannapieco2009} in a suit of eight isolated galaxies embedded inside Milky Way type halos. On the other hand, \cite{Yang2021} analyzed \texttt{EAGLE} and \texttt{TNG100} simulations and found that galactic sizes strongly correlate with the spin parameters of their dark matter halos \citep[see the similar conclusion of][]{Grand2017}.\\

\begin{figure}
\centering
\includegraphics[width=1.\columnwidth]{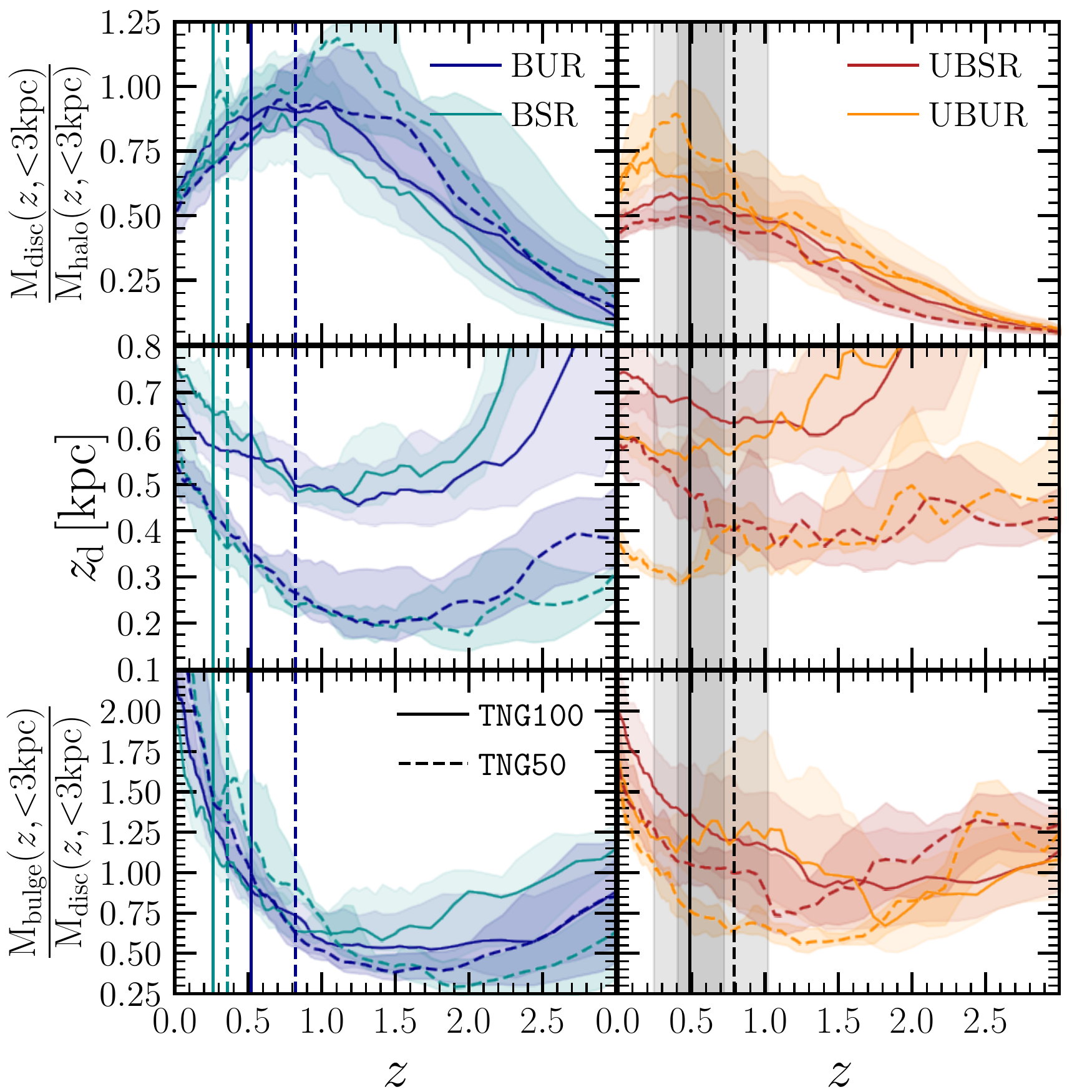}
\caption{Evolution of the median values of the disc-to-halo ($ r\, \rm {<}\,3\,kpc$), vertical scale length of the disc ($z_{\rm d}$) and the bulge-to-disc ($r \,\rm {<}\,3\,kpc$). The color coding, line styles and vertical lines are the same as for Fig.~\ref{fig:Mdisc_Rdisc_Vmax_BUR_BSR_UBUR_UBSR}. }
\label{fig:VSL_B_T_V_MD_MB}
\end{figure}

We also note that unbarred galaxies (UBSR or true negative represents ${\sim}\,80\%$ of the unbarred sample) typically have larger spin values than barred ones. The role of halo spin in the formation and growth of bars has so far been rather controversial.  Using a suite of isolated simulations of discs embedded in dark matter halos with various spin properties, \cite{Saha2013b} showed that the larger the spin of co-rotating halos, the faster, the stronger and the longer is the bar that the simulated galaxy is capable of developing. On the other hand,  \cite{Long2014}, also using a set of isolated galaxy simulations, found that  spinning dark matter halos (${|}{|}\vec{\lambda}_{\rm h}{|}{|}\,{>}\,0.03$) can heavily reduce the secular growth phase of stellar bars and decrease their pattern speed.  The observational study of \cite{Cervantes2013} points towards in a similar direction finding a decrease of the fraction of galaxies hosting strong bars towards large ${|}{|}\vec{\lambda}_{\rm h}{|}{|}$. While it is beyond the scope of this work to study the connection between bar properties and halo spin, our results indicate that most bars (BUR or true positive represents the ${\sim}\,75\%$ of the barred sample) are hosted by more slowly spinning halos, as also shown for the \texttt{TNG50} in \cite{RosasGuevara2021}.\\

In the middle panel of Fig.~\ref{fig:Lh_Concentration_Mh_Ms_BUR_BSR_UBUR_UBSR} we further investigate the dark matter halo properties, looking at the concentration. We  compute it  as the ratio  $\rm V_{max}^{halo}/V_{200}$, where $\rm V_{max}^{halo}$ is the maximum rotational velocity of the halo component. This quantity is a proxy of the DM halo concentration and is widely used in many works \citep[see e.g.][]{Gao2007}, as it does not require any model to fit the simulation data. As shown, \texttt{TNG50} predicts larger concentrations than \texttt{TNG100}. However, small differences are seen between BUR (true positive) and BSR (false negative), and between the UBSR (true negative) and UBUR (false positive) samples. In the \texttt{TNG50} there is a weak indication that barred galaxies (BUR represents the ${\sim}\,75\%$ of the barred sample) are hosted by more concentrated halos, as also discussed by \cite{RosasGuevara2021}. Despite that, the results presented here seem to point out that DM concentration does not play an important role (if any) in the bar formation.\\

To explore the interplay between the DM and stellar components, in the lower panel of Fig.~\ref{fig:Lh_Concentration_Mh_Ms_BUR_BSR_UBUR_UBSR} we present the stellar-to-halo ratio within $3\,\rm kpc$.  As shown in Fig.~\ref{fig:Stellar_Halo_Ratio}, barred galaxies have generally larger stellar fractions than unbarred galaxies. Only for  \texttt{TNG100} the BSR galaxies display slightly smaller stellar-to-halo ratios than  the BUR ones. For \texttt{TNG50}, instead, the stellar-to-halo ratio is larger for the BSR sample. Concerning the unbarred population, we can see that the UBUR and UBSR samples behave in a similar way at $z\,{>}\,1$. However at lower redshifts, their trends diverge, and the UBUR galaxies have larger stellar-to-halo ratios. \\

In Fig.~\ref{fig:VSL_B_T_V_MD_MB} we look at the properties of the disc component. In the top panel, we show the disc-to-halo ratio in the inner region. Both bar samples (BUR and BSR) are characterized by a dominant central disc component, which declines after bar formation as the galaxy develops a pseudobulge or our morphological decomposition classifies some bar particles as a bulge as a consequence of their radial orbits (see the discussion for Fig.~\ref{fig:Stellar_Halo_Ratio}). The UBUR sample has a more dominant disc with respect to the average unbarred population (UBSR), which contributes to the misclassification of these galaxies according to the \ec{}. Still, both unbarred samples have significantly lower disc-to-halo ratios at early times with respect to barred galaxies. Looking at the vertical extent of the disc, we also do not see significant differences between the BUR and BSR samples: all barred galaxies have  generally colder discs with respect to unbarred galaxies, and \texttt{TNG50} produces thinner discs. However, the misclassified unbarred galaxies (UBUR) have thinner discs than the average unbarred population, with values closer to the ones of barred galaxies. Despite this, UBUR galaxies have $z_{\rm d}$ values ${\sim}\,2$ times larger than the ones displayed by BUR and BSR samples. We highlight that the trends at $z\,{\gtrsim}\,1.5$ should be considered with caution since the galaxy morphology is irregular and the resolution of the simulation might have an impact on our results.\\

Finally, in the lower panel of Fig.~\ref{fig:VSL_B_T_V_MD_MB} we explore the evolution of the bulge-to-disc ratio of our sub-samples of barred and unbarred galaxies.  Before bar formation, the BSR sample (false negative) displays slightly larger ratios than the BUR one (true positive), but both barred samples have significantly less predominant bulges with respect to the unbarred sample.  After bar formation, both BUR and BSR galaxies show a fast increase of the bulge component, caused by the development of a pseudobulge structure. The systematically larger $\rm B/D$ ratios of unbarred galaxies suggest that the bulge structure plays an important role in suppressing the development of bars \citep[see, e.g., the early works of  ][]{Ostriker1973,Toomre1981,Sellwood1980,Sellwood2001}. Indeed, using the \texttt{Eris} and \texttt{ErisBH} zoom-in simulations of a Milky Way-type halo, \cite{Bonoli2016} found that early suppression of bulge formation (possibly due to AGN feedback), can lead to discs more prone to instabilities. More recently, using N-body simulations, \cite{Kataria2018} and \cite{Kataria2020} showed a delay in the bar formation as a function of the galaxy bulge-to-disc ratio ($\rm B/D$). Particularly, they reported an upper limit of $\rm B/D \sim 0.2\,{-}\,0.5$ above which the development of a bar is suppressed.\\

Based on the above analysis, we conclude that early disc assembly and the absence of a prominent bulge component are necessary conditions for the formation of a bar. Massive  and compact discs without bars, wrongly classified by the \ec{} as unstable galaxies (UBUR or false positive) have indeed a prominent bulge and assembled at later times. When analyzing the details of the barred samples, the primary difference between the BUR (true positive) and BSR (false negative) galaxies is the average size of the disc, which seems to be linked to the halo spin. In what remains of the section we will further discuss the differences between these two populations, and speculate on the possibility of different trigger mechanisms for their bar structure.\\

\subsection{The missed barred galaxies in the \ec: An external triggering?}

As shown in Section~\ref{sec:ECriterionPerformance}, the BSR sample (false negative) is composed of barred galaxies for which the \ec{} is not able to predict disc instability. As extensively discussed in Section~\ref{sec:ComparingProperties_BUR_UBUR_UBSR_UBUR}, both samples are characterized by early-assembling dominant discs, but the discs of BUR galaxies are significantly more compact, likely because of the lower spin parameter of their host dark matter halos.\\

\begin{figure}
\centering
\includegraphics[width=1.\columnwidth]{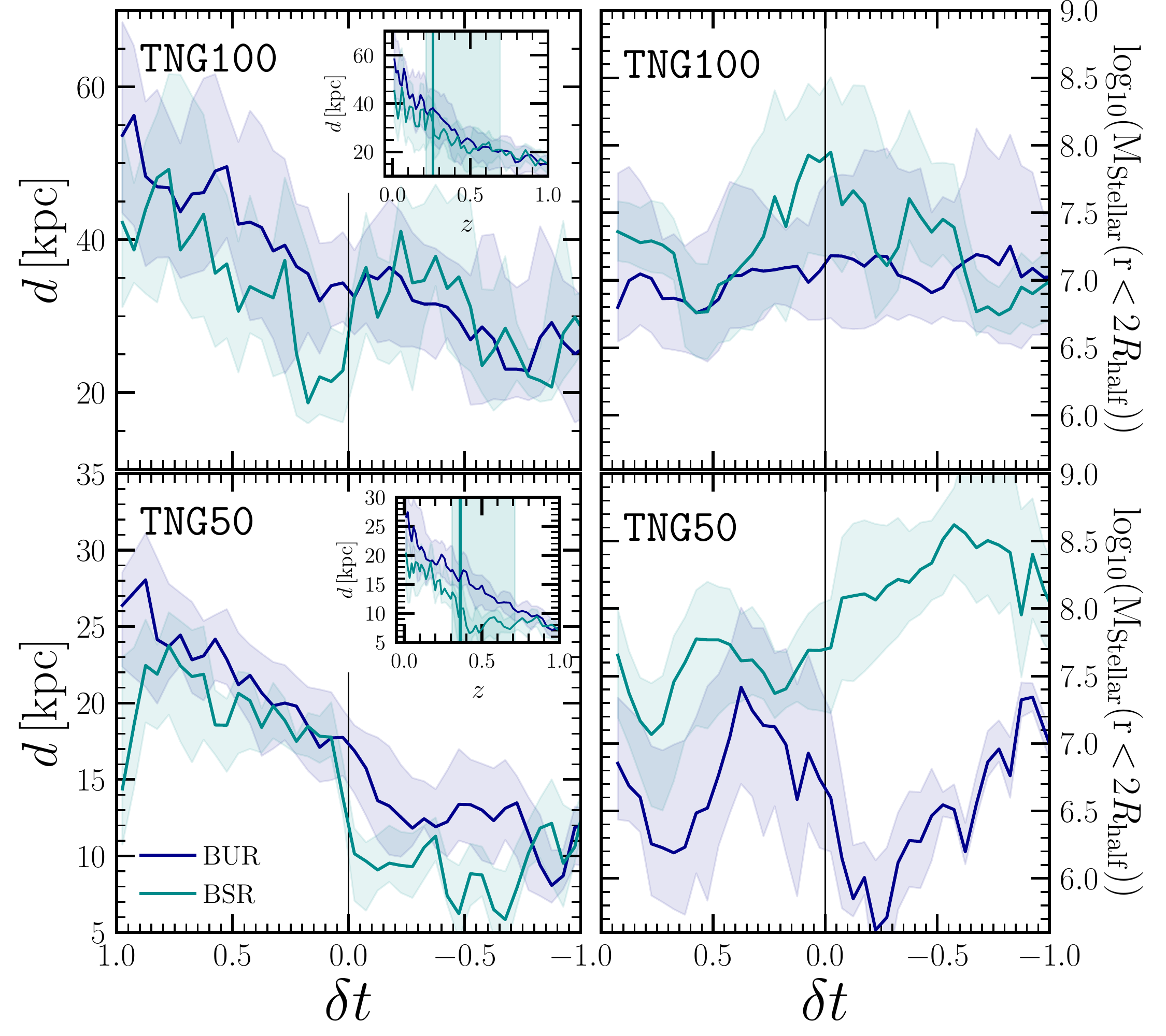}

\caption[]{Distance (left) and stellar mass (right) of the closest neighbour for BUR (dark blue) and BSR (cyan) galaxies as a function of $\delta t$. The dark vertical line highlights $\delta t\,{=}\,0$. The inset displays the same but as a function of redshift for the BUR and BSR samples. The horizontal blue line (shaded area) represents the median ($32^{\rm th}\,{-}\,68^{\rm th}$ percentile) bar formation time of BSR galaxies.}
\label{fig:Distance_To_The_closest_TNG100}
\end{figure}

These differences make us wonder whether the two barred populations (BUR and BSR) have a common origin, or if the instabilities in the two samples could be triggered by different processes.  
In the the general picture of bar formation, which underlies the \ec, disc instabilities are a \textit{secular process}, in which the slow growth of a  self-gravitating stellar disc supported mainly by rotation leads to global dynamical instabilities which then trigger the formation of a bar \citep{Kalnajs1972,Athanassoula1986,Shen2003,Moetazedian2017,Yurin2015,Zana2018A}. However, a number of authors have argued  that bar structures can form (or disappear) during galaxy interactions such as mergers or flybys  \citep[i.e, ``external triggers'', ][]{Miwa1998,Berentzen2004,Lokas2014,PeschkenLokas2019,Ghosh2021}. The  time resolution of the outputs of the \texttt{TNG100} and \texttt{TNG50} simulations is unfortunately not  sufficient to perform a detailed analysis on the role of external triggers on bar formation. Such a study would require a re-running of specific regions of the simulation domain with higher output frequency, as done for instance in \cite{Zana2018A}, and goes beyond the scope of this  paper. In an attempt to assess the role of external triggers in the formation of bars in our samples, in Fig.~\ref{fig:Distance_To_The_closest_TNG100} (left panel) we present the distance to the closest neighbour as a function of time  for the BUR and BSR galaxies. Interestingly, for both  \texttt{TNG100} and \texttt{TNG50}, the distance to the closest neighbour for BSR galaxies drops/experience a fast rise at $-0.15\,{<}\,\delta t\,{<}\,0.15$, i.e around the time of bar formation. In contrast, the distance for the BUR galaxies does not show such a clear feature, with a typical distance values ${\sim}\,1.5\,{-}\,2$ times larger than for BSR. Besides the distance, the stellar mass of the closest neighbour is  also different for BUR and BSR samples (right panel of  Fig.~\ref{fig:Distance_To_The_closest_TNG100}): the neighbours of BSR galaxies are systematically more massive, especially at the bar formation time where the differences can reach up to $1\, \rm dex$.\\ 

Even though this analysis can not conclusively ensure that BSR galaxies developed their bar as a consequence of a close interaction, it provides suggestive clues  in that direction. Indeed, external triggering could explain why the \ec{} is not able to detect the disc instabilities for BSR galaxies. The work of \cite{IzquierdoVillalba2019} used the \texttt{L-Galaxies} SAM to explore the possibility that the \ec{} could detect some disc instabilities caused by external triggering. Specifically, by following the history and the physical conditions of the galaxy in which a disc instability takes place, the authors distinguished between instabilities that are merger-induced and the ones that are a consequence of the slow, secular evolution of galaxies. \cite{IzquierdoVillalba2019} showed that bar/pseudobulge structures in massive galaxies are generally a result of secular processes, although some massive discs can become bar unstable after a minor merger, which prompts the formation of a nuclear ellipsoidal component. In agreement with our findings (see Fig.~\ref{fig:Distance_To_The_closest_TNG100}) these authors reported that merger-induced disc instabilities are rare and occur mostly at $z\,{\sim}\,1$ with a sharp cut-off towards higher redshifts. Despite their attempts to link the \ec{} with an external triggering mechanism, further analysis of this is needed. Indeed, the recent papers of \cite{Zana2018A,Zana2018B} suggest that external perturbers could have a negative effect on the bar formation. By analyzing a state-of-the-art cosmological zoom-in simulations the authors found that minor mergers or close fly-by can delay the bar formation. Even more, they can have a destructive effect, weakening or destroying strong bars. Taking into account the results shown here and all the works presented in the literature, additional dependencies could be added in the \ec{} to determine if a close galaxy encounter or a merger has the capability of triggering/delaying the formation of a bar structure.\\

\section{Conclusions} \label{sec:Conclusions}

In this paper we explored, for the first time in a systematic way, the performance of the \cite{Efstathio1982} analytic criterion (\ec{}) for disc instability using a sample of barred and unbarred galaxies extracted from  large cosmological hydrodynamical simulations. Specifically, we made use of the catalogues of \cite{RosasGuevara2019} and \cite{RosasGuevara2021}, composed of barred and unbarred  disc galaxies (disc-to-total ratios ${>}\,0.5$) extracted from the \texttt{TNG100} and \texttt{TNG50} simulations at $z\,{=}\,0$. To ensure high-enough resolution, we consider only disc galaxies with stellar masses larger than $\rm 10^{10.4}\, M_{\odot}$. Indeed, galaxies with such masses have proven to be the preferential hosts and birthplaces of bar structures in the low-$z$ Universe \citep[see e.g ][]{Gadotti2009,Cervantes2015,Gavazzi2015} \\

We first of all compared the physical properties of strongly barred and unbarred galaxies. By analyzing the disc component we found that the discs of the barred sample generally assembled earlier and the discs are more compact ($\rm {\sim}\,1\, kpc$ of difference). The early assembly of the disc of barred galaxies is related to the early assembly of the dark matter and stellar components, and to early consumption of the available gas. On top of that, the distribution of the baryonic component is different for barred and unbarred galaxies: the central stellar-to-halo   and disc-to-halo ratios are significantly larger for barred galaxies than for unbarred ones, in agreement with other theoretical work \citep{Algorry2017,Fragkoudi2020}. Regarding the bulge component, we found the opposite trend, with barred galaxies being characterized by a subdominant bulge before bar formation. Only after the development of the bar, the bulge component grows, likely because of bar buckling which leads to the formation of a pseudobulge and/or because our morphological decomposition classifies the stellar particles subsiding the bar structure as bulge-like (\textit{hot component}). \\

The differences mentioned above clearly indicate that disc galaxies  that develop a prominent bar structure have generally a very different history than disc galaxies that do not undergo disc instability. To check the ability of the \ec{} to separate the two classes of disc galaxies  to determine the presence or absence of instabilities, we tested its success rate. Effectively, we calculate the fraction of time spent by a galaxy in the bar unstable region according to the \ec{} (i.e., $\epsilon <1.1$). The results showed that the \ec{} is able to detect bar formation in ${\sim}\,75\%$ of the barred galaxies and correctly identifies the absence of a bar structure in $\,{\sim}\,80\%$ of the unbarred galaxies. Despite the large success rate, we still find that the \ec{} fails in ${\sim}\,25\%$ and ${\sim}\,20\%$ of barred and unbarred galaxies, respectively. Carefully analyzing the properties of correctly and wrongly classified galaxies, we reach the following conclusions: 

\begin{itemize}

\item{While some differences exist in the disc mass and maximum circular velocities,  the property in the \ec{} largely responsible for the correct or incorrect classification  is the scale length of the disc: misclassified barred galaxies have a disc which is much more extended than the typical barred population. On the other hand, misclassified unbarred galaxies have small discs, similar to those of average barred galaxies. This seems to be related to the spin parameter of the halo, as we find a positive correlation between the value of the scale length and the spin parameter; less extended galaxies are hosted in more slowly spinning halos. This correlation, theoretically expected \citep[see e.g.,  ][]{Fall1980, MoMaoWhite1997}, was not present in the first simulations of disc galaxies \citep[][]{Scannapieco2009}, but was retrieved in more recent simulation suites \citep{Grand2017}.\\ }

\item {Regarding unbarred galaxies misidentified by the \ec{} as potentially unstable, we found that their overall properties are similar to those of typical barred galaxies: at recent times, they are characterized by massive and compact discs with large stellar-to-halo ratios.  However, the discs assemble later than those of barred galaxies, and the bulge components are significantly more prominent at early times than in barred galaxies. Moreover, the disc vertical scale length is generally larger than in barred galaxies, indicating hotter kinematics. We thus conclude that the bulge-to-disc ratio and/or the disc thickness should also be taken into account when determining disc stability.\\ }


\item{ Concerning the population of barred galaxies that the \ec{} wrongly classifies as stable discs, we found that, they are generally much less compact than typical barred galaxies, and are embedded in dark matter halos with larger spin parameter. Interestingly, we also find that, at the epoch of bar formation, they may often have experienced a close encounter with a massive satellite. This might indicate that, for these galaxies, bar formation could be due to an external trigger, rather than being a consequence of the secular growth of the disc.\\ }
\end{itemize}


Given all the results summarized above, we can conclude that the \cite{Efstathio1982} analytic criterion can robustly describe the stability of most secularly evolving massive disc galaxies. This has important implications for semi-analytic models of galaxy formation (SAMs) since, in most cases, the modelling of disc instability  and  the subsequent growth of the (pseudo)bulge component relies on the \ec{}. Thus, we argue that the predictions of current SAMs regarding massive barred galaxies in the local Universe can be trusted, at least to first order.  Despite this, our analysis suggests that the criterion should be refined to take into account possible externally-induced instabilities or dependencies with the bulge-to-disc ratio. Thanks to the large volume provided by \texttt{TNG100} and the high-resolution of \texttt{TNG50}, we plan to explore in future work different ways to include external triggers for bar formation in the \ec{}, to add extra dependencies that remove contamination and to apply these new criteria in  semi-analytical models.


\section*{DATA AVAILABILITY}

The IllustrisTNG simulations are publicly available and accessible at \href{https://www.tng-project.org/}{https://www.tng-project.org/}. The catalogues of bars can be found in \href{https://www.tng-project.org/data/docs/specifications/}{https://www.tng-project.org/data/docs/specifications/}.

\section*{Acknowledgements}
D.I.V acknowledge financial support provided under the European Union’s H2020 ERC Consolidator Grant ``Binary Massive Black Hole Astrophysics'' (B Massive, Grant Agreement: 818691) and INFN H45J18000450006. S.B. acknowledges partial support from the project PGC2018-097585-B-C22. YRG acknowledges the support of the
“Juan de la Cierva Incorporation” fellowship (ĲC2019-041131-I) and the European Research Council through grant number ERC-StG/716151. AL acknowledges support from MIUR under the grant PRIN 2017- MB8AEZ. The \texttt{IllustrisTNG} simulations were undertaken with compute time awarded by the Gauss Centre for Supercomputing (GCS) under GCS Large-Scale Projects GCS-ILLU and GCS-DWAR on the GCS share of the supercomputer Hazel Hen at the High Performance Computing Center Stuttgart (HLRS), as well as on the machines of the Max Planck Computing and Data Facility (MPCDF) in Garching, Germany.



\bibliographystyle{mnras}
\bibliography{references} 




\appendix

\section{The disc scale length from a kinematics decomposition} \label{appendix:Check_Galaxy_Decomposition}

\begin{figure}
\centering
\includegraphics[width=1.\columnwidth]{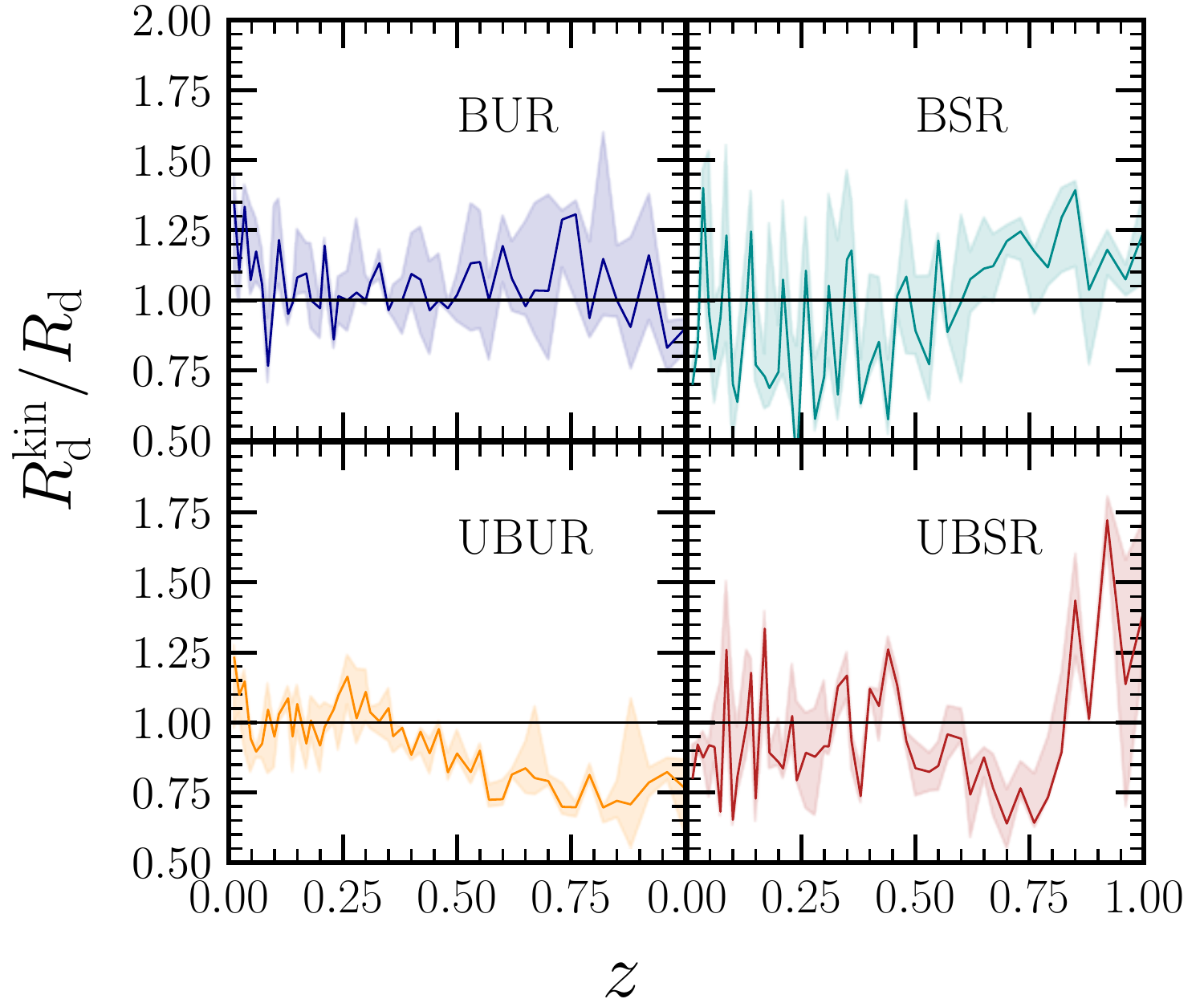}
\caption[]{Ratio between the scale length computed by fitting the mass surface density of thin disc particles ($R_{\rm d}^{\rm kin}$) and the scale length computed according to Section~\ref{sec:Bars_NoBars_and_Fit} ($R_{\rm d}$). Solid line represents the median whereas the shaded area displays the $\rm 32^{th}\,{-}\,68^{th}$ percentile. Dark blue, cyan, orange and red lines are the results for BUR, BSR, UBUR and UBSR \texttt{TNG100} samples.}
\label{fig:Check_Morphology_Zana_This_Wrok}
\end{figure}

In this appendix we compare our values of the disc scale length, $R_{\rm d}$, with the ones computed based on a kinematics decomposition, $R_{\rm d}^{\rm kin}$. In particular, the latter has been computed making use of the morphological classification of Zana et al. (submitted) which, based on the particle kinematics, is capable of distinguishing for any galaxy 5 different components: thin disc, thick disc, bulge, pseudobulge and stellar halo. By selecting only \textit{think disc particles}, we have performed an exponential fit to the resulting face-on surface density profile, $\Sigma_{\rm stars}(r)$ (see second term of Eq.~\ref{eq:Double_sersic_profile}). In Fig.~\ref{fig:Check_Morphology_Zana_This_Wrok} we present the ratio between $R_{\rm d}$ and $R_{\rm d}^{\rm kin}$ for $5$ random galaxies in the BUR, BSR, UBUR and UBSR \texttt{TNG100} sample 
As shown, for all the cases the ratio varies around $1$, with few cases where $R_{\rm d}/R_{\rm d}^{\rm kin}$ is outside the range $0.75$ to $1.25$.

\bsp	
\label{lastpage}
\end{document}